\documentclass[12pt]{article}
\usepackage{amsmath,amssymb,epsfig,amsfonts}
\usepackage{graphicx,subfigure}
\usepackage[usenames, dvipsnames]{color}
\usepackage{tikz}
\usetikzlibrary{3d,calc}
\usepackage[backref]{hyperref}
\usepackage[nosort]{cite}
\usepackage{lscape}
\usepackage{rotating}

\usepackage{extarrows}

\usepackage{verbatim} %for comments

\usepackage[top=1.25in,bottom=1.25in,right=0.85in,left=0.85in]{geometry}

\newcommand{\lefthook}{\lrcorner}

\makeatletter

\newcommand{\ov}[1]{\overline #1}

\newcommand{\bZ}{\mathbb{Z}}
\newcommand{\bC}{\mathbb{C}}
\newcommand{\bP}{\mathbb{P}}
\newcommand{\bR}{\mathbb{R}}

\newcommand{\cN}{\mathcal{N}}
\newcommand{\cM}{\mathcal{M}}

\def\unit{{1\kern-.65ex {\rm l}}}
\def\1{{1\kern-.65ex {\rm l}}}

\newcount\hour \newcount\minute
\hour=\time \divide \hour by 60
\minute=\time
\def\now{%
\ifnum \hour<13
  \ifnum \hour=0 \advance \hour by 12 \number\hour:\else \number\hour:\fi%
     \ifnum \minute<10 0\fi%
     \number\minute%
\ A.M.%
\else \advance \hour by -12 \number\hour:%
  \ifnum \minute<10 0\fi%
  \number\minute%
  \ P.M.%
\fi%
}

\makeatother

\begin{document}

% format
\baselineskip=18pt  % a la harvmac
\numberwithin{equation}{section}  % make eq labels (sec.num)
\allowdisplaybreaks  % allow page breaks in displayed eqs

%%%%%%%%%%%%%%%%%%%%%%%%%%%%%%%%%%%%%%%%%%%
%%%        TITLE BEGINS HERE
%%%%%%%%%%%%%%%%%%%%%%%%%%%%%%%%%%%%%%%%%%%

%% ========== title (note version) begins here ==========
%
%\vspace*{-1cm}
%\begin{center}
% {\Large\bf Title of the Document}
%\end{center}
%\vspace*{-.5cm}
%
%% ========== title (note version) ends here ==========

%% ========== title (paper version, a la harvmac) begins here ==========

\thispagestyle{empty}

% Report number
\vspace*{-3cm}
\begin{flushright}
{\tt NSF-KITP-15-098}

{\tt UCSB Math 2015-10}
\end{flushright}

% title, authors, affiliation
\vspace*{0.8cm}
\begin{center} {\Huge On Gauge Enhancement and Singular Limits \\
    \vspace{.2cm} in $G_2$ Compactifications of
    M-theory% Gauge Enhancement and Topological Defects \\ \vspace{.2cm} in $G_2$ Compactifications of M-theory
  }\\

\vspace*{1.5cm}
James Halverson$^1$ and David R. Morrison$^2$
\vspace*{1.0cm}

$^1$ {\it Kavli Institute for Theoretical Physics, \\
University of California, Santa Barbara, CA 93106, USA}\\
{\tt jim@kitp.ucsb.edu} \\ \vspace{.5cm}

$^2$ {\it Departments of Mathematics and Physics, \\
University of California, Santa Barbara, CA 93106, USA}\\
{\tt drm@physics.ucsb.edu}

\vspace*{0.8cm}
\end{center}
\vspace*{.1cm}

% abstract

\vspace{.5cm} We study the physics of singular limits of $G_2$
compactifications of M-theory, which are necessary to obtain a
compactification with non-abelian gauge symmetry or massless charged
particles. This is more difficult than for Calabi-Yau
compactifications, due to the absence of calibrated two-cycles that
would have allowed for direct control of W-boson masses as a function
of moduli.  Instead, we study the relationship between gauge
enhancement and singular limits in $G_2$ moduli space where an
associative or coassociative submanifold shrinks to zero size; this
involves the physics of topological defects and sometimes gives
indirect control over particle masses, even though they are not
BPS. We show how a lemma of Joyce associates the class of a
three-cycle to any $U(1)$ gauge theory in a smooth $G_2$
compactification. If there is an appropriate associative submanifold
in this class then in the limit of nonabelian gauge symmetry it may be
interpreted as a gauge theory worldvolume and provides the location of
the singularities associated with non-abelian gauge or matter
fields. We identify a number of gauge enhancement scenarios related to
calibrated submanifolds, including Coulomb branches and non-isolated
conifolds, and also study examples that realize them.

%%%%%%%%%%%%%%%%%%%%%%%%%%%%%%%%%%%%%%%%%%%
%%%           TITLE ENDS HERE
%%%%%%%%%%%%%%%%%%%%%%%%%%%%%%%%%%%%%%%%%%%

\clearpage
\tableofcontents
%\printindex

%%%%%%%%%%%%%%%%%%%%%%%%%%%%%%%%%%%%%%%%%%%
%%%        MAIN TEXT BEGINS HERE
%%%%%%%%%%%%%%%%%%%%%%%%%%%%%%%%%%%%%%%%%%%
\newpage

\section{Introduction}

M-theory compactifications to four dimensions are of fundamental
physical interest. Since the weakly coupled superstring theories can
be obtained as limits of M-theory 
\cite{Sen:1994fa,Hull:1994ys,Witten:1995ex}, it is
reasonable to expect that M-theory compactifications give a broad view
of the four-dimensional landscape. However, detailed progress has
historically been limited by both our primitive understanding of
M-theory and also a lack of control over the relevant seven-manifolds,
which are chosen to have $G_2$ holonomy in order to preserve $\cN=1$
supersymmetry in four dimensions \cite{Papadopoulos:1995da}.

Seven-manifolds with $G_2$ holonomy, which we shall simply refer to as
$G_2$ manifolds, are notoriously difficult to construct, in part due
to the lack of an analog of Yau's theorem that might provide a
topological criterion sufficient for the existence of a $G_2$
metric. Additionally, they are real manifolds and therefore cannot be
described directly using the techniques of complex algebraic geometry.
The first compact $G_2$ manifolds were constructed by Joyce
\cite{MR1424428,Joyce2} by resolving singularities on orbifolds;
another construction was subsequently discovered by Kovalev
\cite{KovalevTCS} utilizing ``twisted connected sums.'' In recent years,
there has been much progress\cite{KovalevLee,Corti:2012kd} in using
generalizations of Kovalev's construction to produce large classes of
compact $G_2$ manifolds. For example, \cite{MR3109862} identified
fifty million appropriate ``matching pairs'' of asymptotically
cylindrical Calabi-Yau threefolds that can be used to construct
compact $G_2$ manifolds; perhaps this should be compared to the five
hundred million reflexive polytopes that can be used to construct
Calabi-Yau threefolds. See \cite{Halverson:2014tya} for a recent study
of M-theory on twisted connected sum $G_2$ manifolds.

Encouraged by this significant mathematical progress, we will study
some interesting open questions concerning $G_2$ compactifications of
M-theory. Our aims are twofold. First, as the number of compact
examples has grown we will study the physics of $G_2$
compactifications utilizing objects that are natural in global
compactifications, rather than relying on local descriptions.  Second,
since M-theory on a smooth $G_2$ manifold can realize at most an
abelian gauge sector without massless charged matter, we will study
various limits in moduli space that may give rise to non-abelian gauge
enhancement or massless charged matter. In such limits, the $G_2$
manifold $X$ must develop singularities. The non-abelian gauge group
is determined by the structure of singularities in codimension four
while the non-chiral and chiral matter spectra are determined by
codimension six and seven singularities, respectively. We note that
there is an extensive literature on this topic, particularly in the
local case, including early non-compact work on singularities and
gauge enhancement \cite{Acharya:2000gb,Atiyah:2001qf,Acharya:2001gy},
the relationship to anomaly cancellation \cite{Witten:2001uq},
the study of metrics with $G_2$ holonomy including singular limits
\cite{Cvetic:2001zx,Cvetic:2001ih}, uplifts from chiral type IIa models
with intersecting D6-branes \cite{Cvetic:2001nr,Cvetic:2001kk}, brane
probes \cite{Gukov:2002es} and fibrations \cite{Gukov:2002jv}, the physical
structure of singularities (\cite{Acharya:2004qe} and references
therein), a $G_2$-motivated extension of the MSSM
\cite{Acharya:2008zi} and related phenomenological models
\cite{Acharya:2012tw}.

Regarding gauge enhancement, there are critical differences between
$G_2$ and Calabi-Yau compactifications of M-theory; namely, the method
typically utilized to study gauge enhancement in Calabi-Yau
compactifications cannot be applied directly in the case of $G_2$
compactifications. That method is to choose a smooth manifold in which
the mass of W-bosons is determined by the volume of particular
two-cycles and then to take a limit in the moduli space such that the
two-cycles shrink to zero volume and gauge enhancement occurs. The
method is reliable for Calabi-Yau compactification because the K\"
ahler form calibrates two-cycles, and thus the volumes of holomorphic
curves (which determine W-boson masses) can be reliably computed as
functions of the K\" ahler moduli. Physically, this correlates with
the fact that Calabi-Yau compactifications of M-theory have BPS
particles. In contrast, there is no calibration form for two-cycles in
a $G_2$ manifold and there are no BPS particles in $G_2$
compactifications of M-theory, and therefore one must find other means
to study gauge enhancement. This is the subject of the current
paper.\footnote{For an earlier analysis of another situation in which
  BPS particles do not control gauge enhancement, see
  \cite{confinement}.  In that case, BPS strings provided the
  confining flux tubes emanating from magnetically charged states.
  For Calabi-Yau cases in which gauge enhancement occurs when
  two-cycles collapse due to movement in complex structure moduli,
  rather than K\" ahler moduli, see
  \cite{Grassi:2013kha,Grassi:2014sda,Grassi:2014ffa}.}

Earlier studies of gauge enhancement \cite{Acharya:2000gb,Atiyah:2001qf,Acharya:2001gy} used a weak coupling limit which geometrically corresponds 
to making the volume of the singular locus large; this is difficult to
control in $G_2$ moduli space.
Our approach will be to study gauge enhancement via limits in $G_2$
moduli space in which an associative threefold or coassociative
fourfold shrinks to zero size. These limits are closely related to the
physics of domain walls, instantons, and strings obtained from wrapped
M-branes, but we will also see in some scenarios that they are related
to charged matter. We will show that a lemma of Joyce determines the
class of a three-cycle for any $U(1)$ gauge theory in a smooth $G_2$
compactification and that, if this class is realized by a submanifold
satisfying some additional assumptions, then in the limit of nonabelian
gauge symmetry it is natural to interpret the submanifold
 as the location of the
gauge theory worldvolume. (Instantons and 't Hooft-Polyakov monopoles
play an interesting role in this analysis.) In such a case 
the submanifold provides
a useful guide for gauge enhancement, since the appearance of
singularities in or along it may correspond to the appearance of
massless charged particle states. We will also identify a number of
different scenarios in which it is possible to control charged
particle masses indirectly via calibrated submanifolds, and also
discuss the physics of defects (such as BPS strings) associated with
symmetry breaking.

We begin in section \ref{sec:G2 review} by reviewing elements of
smooth $G_2$ compactifications of M-theory.  In section
\ref{sec:boundary} we discuss the parameter space of $G_2$ manifold
structures (up to diffeomorphism) on a fixed differentiable manifold
and make a natural proposal concerning its boundary.  In section
\ref{sec:topological defects and singular limits} we will study gauge
enhancement broadly in $G_2$ compactifications, discussing associated
topological defects and singular limits. BPS defects can be studied as
functions of moduli since they are obtained by wrapping M2-branes and
M5-branes on calibrated submanifolds; a natural singular limit of $X$
is to tune the volume of these cycles to zero.  Additionally, there is
a moduli-dependent inequality due to Joyce that determines the class
of a non-trivial three-cycle that is useful for thinking about
non-abelian gauge enhancement.  We will use these ideas in section
\ref{sec:gauge enhancement} to study the physics of a number of
different scenarios, including Coulomb branches (associated to the
breaking of a non-abelian group by an adjoint chiral multiplet) and
also conifold transitions. Finally, in section \ref{sec:examples} we
will apply these techniques to concrete examples and see the
appearance of some well-known defects that appear in broken gauge
theories, for example the 't Hooft-Polyakov
monopole\cite{'tHooft:1974qc,Polyakov:1974ek}.

\section{Compactifications of M-theory on $G_2$ Manifolds}
\label{sec:G2 review}
In this section we will review elementary facts about Kaluza-Klein reduction of
eleven dimensional supergravity on seven manifolds with $G_2$ holonomy.

\vspace{.5cm}
\noindent \textbf{\emph{$G_2$ Manifolds}}

\noindent First, let us review some basic facts about $G_2$ manifolds. See
\cite{Corti:2012kd,joyce2000compact} for further details.

 A \emph{$G_2$-structure} on a seven-manifold $X$ is a principal
  subbundle of the frame bundle of $X$ which has structure group
  $G_2$. Practically, each $G_2$ structure gives rise to a three-form
  $\Phi$ and a metric $g_\Phi$ such that every tangent space of $X$ admits
  an isomorphism with $\bR^7$ which identifies $g_\Phi$ with $g_0\equiv
  dx_1^2 + \dots + dx_7^2$ and identifies $\Phi$ with 
  \begin{equation} \Phi_0 =
  dx_{123} + dx_{145} + dx_{167} + dx_{246} - dx_{257} - dx_{347} -
  dx_{356},
  \end{equation}
  where $dx_{ijk} \equiv dx_i \wedge dx_j \wedge dx_k$.  Note that the
  subgroup of $GL(7,\bR)$ which preserves $\Phi_0$ is the
  exceptional compact Lie group $G_2$. The three-form $\Phi$, sometimes called
  the \emph{$G_2$-form}, determines an orientation, a Riemannian
  metric $g_\Phi$, and a Hodge star\footnote{For the four- and eleven-dimensional
  Hodge star, we will use $\star_4$ and $\star_{11}$, respectively.} operator $\star_\Phi$.
%  which we will often shorten to $\star$. 
We will refer to the pair
  $(\Phi,g_\Phi)$ as a $G_2$-structure.

  For a seven-manifold $X$ with a $G_2$-structure $(\Phi,g_\Phi)$
  and associated Levi-Civita connection $\nabla$, the \emph{torsion}
  of the $G_2$-structure is $\nabla \Phi$, and when $\nabla
  \Phi=0$ the $G_2$ structure is said to be
  \emph{torsion-free}. The following are equivalent:
  \begin{itemize}
  \item $Hol(g_\Phi) \subseteq G_2$
  \item $\nabla \Phi = 0$, and
  \item $d\Phi = d \star_\Phi \Phi = 0$.
  \end{itemize}
  The triple $(X,\Phi,g_\Phi)$ is called a \emph{$G_2$-manifold} if
  $(\Phi,g_\Phi)$ is a torsion-free $G_2$-structure on $X$. Then by
  the above equivalence, the metric $g_\Phi$ has $Hol(g_\Phi)\subseteq
  G_2$ and $g_\Phi$ is Ricci-flat. For a compact $G_2$-manifold $X$,
  $Hol(g_\Phi)=G_2$ if and only if $\pi_1(X)$ is finite   
  \cite{Joyce2}.
In this case the
  moduli space of metrics with holonomy $G_2$ is a smooth manifold of
  dimension $b_3(X)$.

We assume for simplicity that the cohomology of $X$ is torsion-free.
Let us set the notation we will use for classical intersection
numbers. We take Poincar\'e dual integral cohomology bases 
$\sigma_I \in H^2(X,\bZ)$ and $\tilde
\sigma^I \in H^5(X,\bZ)$ and corresponding dual homology bases 
$\Sigma^I \in H_2(X,\bZ)$ and
 $\tilde \Sigma_I \in H_5(X,\bZ)$. These satisfy
\begin{equation}
  \int_X \sigma_I \wedge \tilde \sigma^J = \tilde \Sigma_I \cdot \Sigma^J = \delta^J_I
= \sigma_I(\Sigma^J) = \tilde \sigma^J(\tilde \Sigma_I).
\label{eq:two five basis duality}
\end{equation}
Similarly, we take Poincar\'e dual integral cohomology
bases $\Phi_i \in H^3(X,\bZ)$ and
$\tilde \Phi^i\in H^4(X,\bZ)$, along with the corresponding dual homology bases
$T^i\in H_3(X,\bZ)$ and $\tilde T_i \in H_4(X,\bZ)$. These satisfy
\begin{equation}
\int_X \Phi_i \wedge \tilde \Phi^j = \tilde T_i \cdot T^j = \delta_i^j
= \Phi_i(T^j) = \tilde \Phi^j(\tilde T_i).
\label{eq:three four basis duality}
\end{equation}
We use capital Latin indices for two-forms,
five-forms, two-cycles, and five-cycles; similarly, we use lower
case Latin indices for three-forms, four-forms, three-cycles, and
four-cycles.
The classical intersection numbers in a simply-connected $G_2$ manifold are
\begin{equation}
  C_{IJk} \equiv \int_X \sigma_I \wedge \sigma_J \wedge \Phi_k = \tilde \Sigma_I \cdot \tilde \Sigma_J \cdot \tilde T_k
\label{eq: classical intersections}
\end{equation}
and, as one would expect, these numbers play a critical role in determining the
physics of the four-dimensional $\cN=1$ compactification. 

Since  $X$ is a manifold with $G_2$ structure, vectors and differential
forms on $X$ fall into $G_2$-representations. The decomposition of forms
into $G_2$ representations extends to cohomology \cite{BryantMetrics1987}: 
\begin{align}
  H^2(X,\bR) &= H_7^2 \oplus H_{14}^2 \nonumber \\
  H^3(X,\bR) &= H_1^3 \oplus H_{7}^3 \oplus H_{27}^3
  \end{align}
  where the summands are defined as\footnote{Since this section is
devoted to descriptions in cohomology rather than specific forms, 
we suppress the dependence of
$\star$ on the $G_2$ form.}
  \begin{align}
  H_7^2 &= \{ \star(\alpha \wedge \star \Phi) \,\, | \,\, \alpha \in H^1(X,\bR) \} \nonumber \\
        &= \{\sigma \in H^2(X,\bR) \,\, | \,\, \star(\sigma \wedge \phi) = 2 \sigma \} \nonumber \\ 
  H_{14}^2  &= \{\sigma \in H^2(X,\bR) \,\, | \,\, \star(\sigma \wedge \phi) = - \sigma \}  \nonumber \\
  H_{1}^3 &= \{ r \, \Phi \,\, | \,\, r \in \bR \} \nonumber \\ 
  H_7^3 &= \{ \star(\alpha \wedge \Phi) \,\, | \,\, \alpha \in H^1(X,\bR)\} \nonumber \\ 
  H_{27}^3 &= \{\alpha \in H^3(X,\bR) \,\, | \,\, \alpha \wedge \Phi = 0 \,\,\, \text{and}\,\,\, \alpha \wedge \star \Phi = 0 \}.
\end{align}
It will be important for us that if $X$ has holonomy exactly $G_2$,
instead of simply being a manifold with $G_2$ structure, then $H_7^2$ is empty
and therefore $H^2(X,\bR)=H_{14}^2.$

This last observation is useful for the following reason. Consider any harmonic two-form
$\sigma$. Then we have \cite{BryantMetrics1987}
\begin{equation}
\star_\Phi \sigma= - \sigma \wedge \Phi.
\label{eq:five-form equation}
\end{equation}
This is a nice relation; the Hodge star $\star_\Phi$ explicitly
depends on the metric, but this dependence is encoded in a simple way
in the metric moduli. Therefore, anywhere the Hodge star of a two-form
appears in a Kaluza-Klein reduction, the moduli dependence implicit in
$\star_\Phi$ can be encoded directly in terms of $\Phi$. For example for
harmonic two-forms $\sigma^1$ and $\sigma^2$,
\begin{equation}
  \int \sigma^1 \wedge \star_\Phi \sigma^2 = - \int \sigma^1 \wedge \sigma^2 \wedge \Phi = - \int (s^1_I\, \sigma_I) \wedge (s^2_J\, \wedge \sigma_J) \wedge (\phi_k \, \Phi_k) = 
  - s^1_I s^2_J \phi_k\, C_{IJk}.
\end{equation}
Expressions of this form determine the structure of abelian gauge
coupling functions and kinetic mixings obtained from Kaluza-Klein reduction, for example.

\vspace{.5cm}
\noindent \textbf{\emph{Calibrated Geometry for $G_2$ Manifolds}} 

In both the Calabi-Yau and $G_2$ manifold cases, we lack explicit knowledge
of the metric.  However, in both cases,
the volumes of certain
cycles in this metric can be computed via calibrated geometry as
developed in the seminal work of Harvey and Lawson
\cite{HarveyLawson}. Their fundamental observation is the
following. Let $X$ be a Riemannian manifold and $\alpha$ a closed
$p$-form such that $\alpha|_\xi \leq vol_\xi$ for all oriented tangent
$p$-planes $\xi$ on $X$. Then any compact oriented $p$-dimensional
submanifold $T$ of $X$ with the property that $\alpha|_T=vol_T$ is
a minimum volume representative of its homology class, that is
\begin{equation}
vol(T) = \int_T \alpha = \int_{T'} \alpha \leq vol(T')
\end{equation}
for any $T'$ such that $[T-T']=0$ in $H_p(X,\bR)$. Note in particular
the useful fact that $vol(T)$ is computed precisely by $\int_T
\alpha$, even though one may not know the metric on $X$ explicitly.

If $X$ is a Calabi-Yau threefold, the K\" ahler form $J$, the
holomorphic three-form $\Omega$, and the square of the K\"ahler form $J\wedge J$
 are calibration forms for two-cycles,
three-cycles, and four-cycles; they calibrate holomorphic curves,  special
Lagrangian submanifolds, and holomorphic divisors. 
Note that, in M-theory compactifications
on $X$ the presence of calibrated two-cycles allows for control over
massive charged particle states obtained from wrapped M2-branes 
\cite{Witten:1996qb}. This
computes particle masses as a function of moduli.

If $X$ is a $G_2$ manifold, $\Phi$ and $\star_\Phi \Phi$ are calibration forms
which calibrate so-called associative three-cycles and coassociative
four-cycles, respectively. As we will discuss, this allows for control over
topological defects obtained from wrapping M2-branes and M5-branes on
calibrated  three-cycles and four-cycles; these are instantons, domain
walls, and strings. Note the absence of calibrated two-cycles, however.

\vspace{.5cm}
\noindent \textbf{\emph{Elements of Kaluza-Klein Reduction}}

While we do not aim to be exhaustive, we would like to briefly review
certain aspects of the Kaluza-Klein reduction of eleven dimensional
supergravity on $X$, pioneered in \cite{Papadopoulos:1995da}.
We ignore the $\alpha'$ corrections, which have been considered recently
\cite{Becker:2014rea,Becker:2014uya,Becker:2015cca}.

Consider a compactification of M-theory on a smooth $G_2$ manifold $X$
at large volume, where the metric of $X$ is determined by a
torsion-free $G_2$ form $\Phi \in H^3(X)$, and $\Psi \equiv \star_\Phi
\Phi$ is the dual four-form. This gives rise to a four-dimensional
$\cN=1$ effective theory, which can be obtained via Kaluza-Klein
reduction of 11-dimensional supergravity. 
In order to perform this
reduction, we expand the M-theory three-form $C_3$ as
\begin{equation}
\label{eqn:C3 ansatz}
  C_3 = A_I\wedge \sigma_I + \theta_i\wedge \Phi_i, 
\end{equation}
where the $A_I$ are four-dimensional abelian vector potentials and the
$\theta_i$ are pseudo-scalars in four-dimensions. The latter, together
with the scalars obtained from reduction of metric moduli via the
$G_2$-form $\Phi$, form complex scalars that sit in chiral multiplets.
Similarly, the dual six-form can be expanded as
\begin{equation}
\label{eqn:C6 ansatz}
  C_6 = \tilde A^I \wedge \tilde \sigma^I + B^i \wedge \tilde \Phi^i,
\end{equation}
where $\tilde A^I$ are the dual magnetic vector potentials upon
reduction to four dimensions and the $B^i$ are two-forms which are
four-dimensional Hodge duals of the $\theta$, i.e. $d\theta =
\star_4\, dB$. We call them $B^i$ rather than $\tilde \theta^i$ to
align conventions with the weakly coupled type II superstring
literature, where such two-forms $B$ give rise to the $\int d^4x\, B
\wedge F$ type St\"uckelberg couplings critical for anomaly
cancellation via the Green-Schwarz mechanism. In summary, the supergravity
sector of M-theory on $X$ gives rise to $b_3(X)$ neutral chiral multiplets and
$b_2(X)$ $U(1)$ vector multiplets.

The bosonic part of the eleven-dimensional supergravity action is given by
\begin{equation}
  \label{eq:11dsugrabosonic}
  S = \frac{1}{2\kappa_{11}^2}\left(\int d^{11}x \,\, \sqrt{-g}\, (R-\frac{1}{2}|G_4|^2) - \frac{1}{6}\int C_3 \wedge G_4 \wedge G_4\right) 
\end{equation}
where $G_4$ is the field strength of the M-theory three-form. The
Chern-Simons term gives rise to abelian $\theta$-terms in the
four-dimensional action
\begin{equation}
S_{4d,\text{theta}} = -\frac{1}{12 \kappa_{11}^2} C_{iJK} \int  \theta_i\, F_J \wedge F_K.
\end{equation}
The kinetic term for the M-theory three-form gives rise to kinetic
terms for the four-dimensional gauge fields. Generically in a
four-dimensional $\cN=1$ supergravity theory these are of the form
\begin{equation}
  S_{4d,\text{kinetic}} = -\frac{1}{4} Re\, f_{JK}(\Phi)\, \int F_J \wedge \star_4 F_K
\end{equation}
where the prefactor $f_{JK}$ is the gauge coupling function. These kinetic terms arise from the $\int G_4
\wedge \star_{11} G_4$ terms in \eqref{eq:11dsugrabosonic} and their dependence
on $G_2$ takes the form $\int_X \sigma_J \wedge \star_\Phi \sigma_K$ which can be rewritten
in terms of $G_2$ moduli using the property $\star_\Phi \sigma = - \sigma \wedge \Phi$
characteristic of $H_{14}^2(X,\bZ)$. We will use this moduli dependence in the following.

Alternatively, a different Lagrangian for eleven-dimensional supergravity arises if different
mass dimensions are chosen for the three-form field in the theory. Following \cite{Svrcek:2006yi}, taking $2\kappa_{11}^2=l_{11}^9/2\pi$,
$\tilde C_3 = C_3/l_{11}^3 = C_3/(4\pi\kappa_{11}^2)^{1/3}$, and associated field strength $\tilde G_4$
the action is
\begin{equation}
S_{11}=2\pi \left(\frac{1}{l_{11}^9}\int d^{11}x\,\,\sqrt{-g} R-\frac{1}{2l_{11}^3}\int d^{11}x \,\, |\tilde G_4|^2 - \frac{1}{6}\int d^{11}x \,\,\tilde C_3\wedge \tilde G_4\wedge \tilde G_4\right).
\end{equation}
Of course, changing the mass dimensions in this way affects the induced mass dimensions of  four-dimensional fields upon compactification, and therefore also
the mass dimensions of the four-dimensional coupling constants. We note this alternate form of the action since it gives rise to canonical
mass dimensions for four-dimensional coupling constants, such as a dimensionless gauge coupling.

\section{The Boundary of the Moduli Space}
\label{sec:boundary}

In this section we will review what is known about the boundary of
the K\"ahler moduli space of Calabi--Yau threefolds, and point out analogies
with the moduli space of $G_2$ metrics on a fixed differentiable
manifold.  We identify limits in the moduli space that necessarily
lead to singularities, and speculate that all such limits take this
general form.  In the case of Calabi--Yau threefolds, thanks to the
theorems of Calabi \cite{Calabi} and Yau \cite{Yau}, the K\"ahler moduli space can be identified
with a subset of the second cohomology group, which in fact forms a
cone.

According to Wilson \cite{Wilson}, the K\"ahler cone of a Calabi--Yau
threefold $Z$ is completely determined by three conditions on the
K\"ahler form $\omega$:
\begin{align}
& \text{(1) a topological condition: } \int_Z \omega\wedge\omega\wedge\omega > 0, \\
& \text{(2) a characteristic class condition: } \int_Z p_1(Z)\wedge\omega <0,
\text{ and} \\
& \text{(3) two calibrated cycle conditions: }  \notag \\
&\qquad \text{for every effective
algebraic curve } C, \quad \int_C \omega>0, \text{ and } \\
&\qquad \text{for every effective algebraic surface } S,
 \quad \int_S \star \, \omega >0 . 
\end{align}
The characteristic class condition \cite{miyaoka-chern} is usually
stated in terms of $c_2(Z)=-\frac12p_1(Z)$ rather than $p_1(Z)$, and
the condition on surfaces, which is normally written in the form
$\int_S \omega\wedge \omega > 0$ \cite{nakai,moishezon}, is usually
omitted entirely since it follows from the condition on curves
\cite{kleiman-ampleness}.

As Wilson further explains, 
except for the theoretical possibility of a curved part of the boundary
of the cone (for which no examples are known), the boundaries of the cone
either correspond to curves or divisors that collapse
to lower dimension as the boundary is approached, or correspond to
fibrations by elliptic curves or K3 surfaces where the fibers collapse
at the boundary.
When the K\"ahler cone is complexified, the codimension one boundary
components become divisors on the boundary of the moduli space, 
and the complexified cone itself becomes a neighborhood of an point of
intersection of such divisors \cite{compact}.

By mirror symmetry, a similar structure holds near large complex
structure limit points in the complex structure moduli space.
Such a point will be the intersection of divisors along which
singularities are acquired via variation of complex structure;
for such singularities, there are always ``vanishing cycles,''
which are $3$-cycles whose volume goes to zero at the divisor.
Although we do not have mathematical results which guarantee
that those $3$-cycles will have special Lagrangian representatives,
by mirror symmetry, they must (since they are mirrors of calibrated
cycles, i.e., the algebraic cycles on the mirror which collapse).
We thus see a very similar structure for complex moduli.

We propose here an analogous structure for the parameter space
of $G_2$ manifold structures up to diffeomorphism-isotopic-to-the-identity
on a fixed differentiable manifold
$X$.  By a result of Joyce \cite[Theorem C]{MR1424428},
by associating the cohomology class $[\Phi]$ to each $G_2$-manifold
structure, we can identify this parameter space with an open subset of
$H^3(X,\mathbb{R})$.

According to Joyce \cite[Lemma 1.1.2]{Joyce2} and
Harvey-Lawson \cite{HarveyLawson},
there are three conditions which such
a the cohomology class of a
$G_2$ form $\Phi$ must satisfy: \phantom{\cite{MR3110581}}
\begin{align}
& \text{(1) a topological condition: } \int_X \sigma\wedge\sigma\wedge\Phi <0 
\text{ for every } \sigma\in H^2(X,\mathbb{R}), \\
& \text{(2) a characteristic class condition: } \int_X p_1(X)\wedge\Phi <0,
\text{ or more generally,\footnotemark\ for any bundle } \\
& \hphantom{\text{(2) }} \text{$E$ admitting a ``$G_2$-instanton''
\cite{Corrigan:1982th,doth}, }
\int_X p_1(E)\wedge\Phi <0,
\text{ and} \label{eq:G2-inst}\\
& \text{(3) two calibrated cycle conditions:\footnotemark }  \notag \\
&\qquad \text{for every 
associative cycle } A, \quad \int_A \Phi>0 , \text{ and} \\
&\qquad \text{for every 
coassociative cycle } C, \quad \int_C \star_\Phi\, \Phi >0 . 
\end{align}
\footnotetext[3]{We thank Thomas Walpuski for bringing
this condition to our attention.  The fact that the Levi--Civita
connection is a $G_2$-instanton on the tangent bundle of $X$
is pointed out in Example 3.4 of \cite{MR3110581}; this
is why $p_1(X)=p_1(TX)$ occurs as a special case.}
\footnotetext{There is a potential problem with this formulation:  the
set of cycles with associative or coassociative representatives may
depend on $\Phi$.  Thus, some of these conditions may not actually generate
boundary walls, since by the time the integral vanishes, the cycle may
no longer be associative and the condition may no longer apply.  
We thank Thomas Walpuski for a discussion on this point.}
Unlike the case of the K\"ahler cone, however, there is no
evidence that these conditions determine the parameter space completely..
 However, since no other phenomena are known which
lead to constraints on the $G_2$ parameter space, we propose that
these known conditions should be a complete list.
In particular, in this paper we shall attempt to study limits of
smooth $G_2$ manifolds by studying the approach to boundary components
which are  defined by the conditions above.

Under appropriate assumptions about the behavior of the volume and
diameter of the metric as the boundary of the parameter space is approached, the
Gromov--Hausdorff theory of limiting Riemannian metrics
\cite{gromov-hausdorff} can be used to show that the singular limit
has the structure of a $G_2$ manifold away from a subset of real
codimension $4$ \cite{cheeger-tian}, and that the generic
singularities in real codimension $4$ are orbifold singularities of
the type which lead to non-abelian gauge symmetry in the
compactification of M-theory.  Any further singularities are in
codimension $6$ or codimension $7$, and in fact the physics of
isolated codimension $7$ singularities has been studied extensively.
In particular, for the examples which have been analyzed in detail
\cite{Atiyah:2001qf}, there is always a vanishing
$3$-cycle\footnote{As pointed out in \cite{Atiyah:2001qf}, in the case
  of vanishing $3$-cycles there can be membrane instantons which
  affect the $C$-field superpartner of $\Phi$, and cause the quantum
  moduli space to differ from the classical moduli space.}  or a
vanishing $4$-cycle (associative, or coassociative, respectively).
This is the phenomenon we are proposing for the general case.

\section{Gauge Enhancement, Defects, and Singular Limits}
\label{sec:topological defects and singular limits}
In this section we will study gauge enhancement in $G_2$ compactifications
of M-theory and its relationship to topological defects and singular limits
of $X$. We will first present the logic for studying gauge enhancement via
singular limits in which calibrated submanifolds vanish.
After presenting the logic, we will study
topological defects and singular limits in detail, and then will
conclude the section with some further comments on gauge enhancement.

Before proceeding, we would like to re-emphasize a main point of our
work discussed in the introduction: understanding and controlling
gauge enhancement is more difficult in $G_2$ compactifications
than in Calabi-Yau compactifications, since (unlike Calabi-Yau manifolds)
there is no calibration form for two-cycles, and therefore it is
more difficult to control W-boson masses as a function of moduli. 
Therefore a controlled description of non-abelian gauge enhancement
requires either gaining some mathematical control over two-cycles
or utilizing different physical techniques; we will do both, though
two-cycles will only be controlled indirectly.

\subsection{Logic Behind Gauge Enhancement in $G_2$ Compactifications}
\label{sec:GE logic}

We use the following logic as a guide for obtaining either non-abelian
gauge enhancement or the existence of massless charged particles in
limits of $G_2$ compactifications.

Since compactification of M-theory on a smooth $G_2$ manifold gives
rise to an abelian theory with no massless charged fields, the
existence of non-abelian gauge symmetry or massless charged fields in
a these compactifications requires taking a singular limit in the
metric moduli space. Conversely, if the singular limit gives rise
to non-abelian gauge symmetry, the process of returning to a smooth
$G_2$ manifold from the singular limit\footnote{It is possible that
if singularities of higher codimension are also present, they might
obstruct any ``return to a smooth $G_2$ manifold.''} necessarily corresponds to
spontaneous symmetry breaking via the Higgs mechanism. The $D$-flat
direction associated to the breaking has a corresponding massless scalar
fluctuation from the broken vacuum, represented by a supergravity mode; as
such, $b_3(X)$ necessarily increases in the Higgsing process. Since the
metric on the $G_2$ manifold $X$ is determined by the $G_2$-form
$\Phi$, taking a singular limit of $X$ requires performing a
deformation $\Phi \mapsto \Phi + \delta \Phi$, which also determines a
deformation of the associated four-form $\star_\Phi \Phi \mapsto 
\star_{\Phi + \delta\Phi}( \Phi
+  \delta \Phi)$.

How might one detect such a singularity, necessary for the existence
of non-abelian gauge symmetry or massless charged matter, after
performing a $\Phi$-deformation?  Unlike in complex algebraic
geometry, $G_2$ manifolds do not admit the luxury of detecting
singularities via the structure of algebraic equations; furthermore,
though $\Phi$ determines a metric, its form is not known in general
compact examples. Given
current knowledge, as explained in the previous section,
it seems to us that the most natural way to do so
is to perform a $\Phi$-deformation of $X$ which shrinks an associative
and / or coassociative submanifold to zero size. Certainly this is
sufficient to produce a singularity.

We therefore propose studying the physics of gauge enhancement via its
relation to vanishing associative or coassociative submanifolds.  This
includes the physics of M2-branes and M5-branes wrapped on such
calibrated submanifolds, which give rise to instantons, strings, and
domain walls. These defects can be controlled because they wrap
calibrated submanifolds in the geometry: they are BPS defects. We will
see the appearance of other topological defects, such as 't
Hooft-Polyakov monopoles, but these cannot be controlled by calibrated
geometry, as expected since there are no BPS particles in $d=4$
$\cN=1$ gauge theories. They will still be useful, though, as we will
see in section \ref{sec: adjoint gauge enhancement} and the first
example of section \ref{sec:examples Joyce}.

\subsection{Defects from Wrapped Membranes and Five-branes}
\label{sec:defects}

In addition to contributions from Kaluza-Klein reduction of
eleven-dimensional supergravity, the physics of M-theory
compactifications depend also on the physics of wrapped M2-branes and
M5-branes. Depending on the details of the wrapped cycles in $X$, a
variety of objects can appear in the non-compact spacetime. These
include not only ordinary particles, but also a variety of topological
defects such as instantons, monopoles, strings, and domain walls. In
some cases these objects that arise in M-theory compactifications can
be identified with objects known to exist in spontaneously broken
gauge theories, and in some cases we will see their parametric
dependence on the expectation values of scalar fields via the
structure of the geometry.

\subsubsection*{Charged Particles and Monopoles from Wrapped Branes
  and Topological Protection}
\label{sec:particles and monopoles}

Let us begin by considering particle states arising from wrapped
branes. Suppose that $X$ has\footnote{Recall that unlike for
  Calabi-Yau manifolds, a $G_2$ manifold may have $b_2(X) = 0$.}
non-trivial positive two-cycles $\Sigma_I$, $I\in \{1,\dots,b_2(X)\}$; then
since $X$ is a smooth manifold, it also has dual five-cycles $\tilde
\Sigma_I$, and if these cycles are submanifolds then particles in
four-dimensions arise from wrapping M2-branes and M5-branes on
$\Sigma_I$ and $\tilde \Sigma_I$, respectively. 

There are two important physical consequences of the fact
that these particles arise from wrapping branes on non-trivial cycles
$\Sigma_I$ and $\tilde \Sigma_I$. The first is related to intersection
theory.  To see this, recall that the Kaluza-Klein ans\"atz for $C_3$
and also $C_6$ takes the form
\begin{equation}
  C_3 = A_I \wedge \sigma_I + \cdots \qquad \qquad \text{and} \qquad \qquad C_6 = \tilde A^I \wedge \tilde \sigma^I + \cdots,
\end{equation}
where $A$ and $\tilde A$ are electric and magnetic vector
potentials in four dimensions. 
An M2-brane (for example) on $L\times \Sigma$ where
$\Sigma \equiv n^I \Sigma^I$ and $L$ is the worldline of the spacetime
particle couples as
\begin{equation}
\int_{L\times \Sigma} C_3 = \sum_I n^I \int_{L \times \Sigma^I} A_J \wedge \sigma_J = n^I \int_L A_I
,
\end{equation}
by virtue of (\ref{eq:two five basis duality}).  This is the
coupling of a particle with charge $n^I$ under the $I^\text{th}$
$U(1)$ associated to the vector potential $A_I$.  These massive
particles could be electrons or W-bosons; it would be interesting
to derive their spacetime quantum numbers explicitly.\footnote{In
M-theory on Calabi-Yau
manifolds, the corresponding particles are BPS and
this fact is used in deriving the spacetime quantum
numbers  \cite{Witten:1996qb,4d-transitions}.}
The topologically non-trivial two-cycles are related to the existence
of geometric\footnote{Here we study $U(1)$'s at the level of geometry,
  and currently have nothing to say about whether other couplings in
  the theory may give them a mass.} $U(1)$ symmetries, and
the topology determines an observable in the four-dimensional theory, namely,
the charge of the particle.

The second physical point is related to stability: an M2-brane wrapped
on a non-trivial volume-minimizing two-cycle $\Sigma$ could be stable
against decay due to charge conservation in the low-energy theory;
this charge conservation would arise from topological protection. This
should be contrasted with a vacuum where the gauge symmetry is
completely broken. There, the W-bosons are not charged, therefore not
protected, and have significantly different physics; this correlates
with the disappearance of non-trivial two-forms after the abelian theory
is Higgsed. If the corresponding two-cycles are still present in the geometry,
they must be trivial in homology.

Wrapping M5-branes on 5-cycles can also give rise to particles; doing so on $L
\times \tilde \Sigma$, where $\tilde \Sigma = n_I\tilde \Sigma_I$ and
$L$ is its worldline, we see that it couples as
\begin{equation}
  \int_{L \times \tilde \Sigma} C_6 = \int_{L\times n_I\tilde \Sigma_I} \tilde A^J \wedge \tilde \sigma^J = n_I\int_L \tilde A^I
\end{equation}
by virtue of (\ref{eq:two five basis duality}), where $n_I$ is the
charge of the particle under the $I^\text{th}$ \emph{magnetic} $U(1)$
vector potential $\tilde A^I$. This is a magnetic monopole. If the
$U(1)$'s of the vacuum state obtained from M-theory on $X$ arise from
the spontaneous breaking of a non-abelian gauge theory, and new
non-trivial five-manifolds appear in $X$ as a result of that breaking,
then these may be the standard monopoles of `t Hooft
\cite{'tHooft:1974qc} and Polyakov \cite{Polyakov:1974ek}. The monopole mass is given by
\begin{equation}
m^2 = 16\pi^2\frac{|v|^2}{g^2}
\label{sec:monopole mass scaling}
\end{equation}
where $v$ is the Higgs field expectation value and $g$ is the gauge
coupling. As we will see in sections \ref{sec:gauge enhancement} and
\ref{sec:examples}, this $|v|^2/g^2$ dependence can be understood
geometrically.

\begin{table}
\centering
\begin{tabular}{c|c|c|c|}
  Submanifold & Dimension & M2 & M5 \\ \hline
  $\Sigma$ & 2 & charged particle & spacetime filling brane \\ \hline
  $T$ & 3 & spacetime instanton & domain wall \\ \hline
  $\tilde T$ & 4 & --- & axionic string \\ \hline
  $\tilde \Sigma$ & 5 & --- & magnetic monopole \\ \hline
\end{tabular}
\caption{$G_2$ manifolds can have non-trivial two-cycles, three-cycles, four-cycles, and five-cycles. By wrapping
  M2-branes and M5-branes on cycles that are submanifolds, a variety of defects can arise in four-dimensions; only those associated to 
  calibrated three-cycle and four-cycles are BPS defects.
}
\label{table:general topological defects}
\end{table}

\subsubsection*{BPS Defects: Instantons, Strings, and Domain Walls}
\label{sec:controllable defects}
While the charged particles (both electrons and monopoles) discussed
in the previous section are difficult to control geometrically,
calibration of three-cycles and four-cycles in $G_2$ manifolds allows
for control over the associated topological defects, which \emph{are}\/
BPS. These are instantons, strings, and domain walls from M2-branes on
three-cycles, M5-branes on four-cycles, and M5-branes on three-cycles,
respectively. See Table \ref{table:general topological defects} for a
listing of the possibilities.

Let us begin by discussing a spacetime instanton obtained by wrapping
an M2-brane on an associative submanifold $T\equiv n^i T^i$. The
classical instanton action depends on the volume
\begin{equation}
  vol(T) = \int_T \Phi = \int_{n^i\,T^i} \phi_j \Phi_j = n^i \phi_i
\end{equation}
and also its coupling to the M-theory three-form
\begin{equation}
  \int_T C_3 = \int_{n^i T^i} \theta_j \Phi_j = n^i \theta_i. 
\end{equation}
That is, the instanton couples to the chiral supermultiplet in
four-dimensions with associated complex scalar field $n^j(\phi_j + i
\theta_j)$.  In the instanton background, the four-dimensional
effective action receives corrections
\begin{equation}
\Delta S_{4d} = \int [dZ]\, A\, e^{-S_{inst}} = \int [dZ] A\, e^{-n^i\,(\phi_i + i \theta_i)+\dots}
\end{equation}
where $[dZ]$ represents integration over the instanton zero modes and
we note the exponential dependence on the four-dimensional complex
scalars $\phi_i + i \theta_i$.  The instanton prefactor $A$ has been studied
in \cite{Harvey:1999as}, though currently more can be said in Calabi-Yau compactifications
of M-theory and also in F-theory; see e.g.
  \cite{Cvetic:2012ts}. 

  The precise nature of the instanton correction depends critically on
  its zero modes; for example if there is an identification $[dZ] =
  d^4x \, d^2\theta$ of the instanton zero modes with the spacetime
  coordinates $x^\mu$ and superspace spinor $\theta_\alpha$ of the
  $d=4$ $\cN=1$ theory, then the instanton correction is a
  superpotential correction. These $(x,\theta)$ zero modes are the
  Goldstone bosons and Goldstinos of the spacetime translations and
  supersymmetries broken by the instanton. Note that, while one might
  assert the irrelevance of so-called $\ov \tau$ instanton zero modes
  in M-theory compactifications, since (unlike in type II compactifications)
  the geometric background only preserves four supercharges, it is known
  that in passing from type IIb to F-theory these modes are repackaged
  \cite{Blumenhagen:2010ja,Cvetic:2010rq} and there is a condition that must
  be checked to ensure their absence. It would be interesting to understand 
  whether a similar repackaging occurs in the IIa to M-theory on $G_2$ limit.

  In addition to the zero modes associated to the super-Poincar\' e
  invariance broken by the instanton, there might also be deformation
  zero modes.  These were some of the zero modes studied by Harvey and
  Moore \cite{Harvey:1999as}, who showed that an M2-brane instanton on
  $T$ corrects the superpotential if $T$ is rigid and supersymmetric,
  i.e. rigid and associative. Together these ensure the absence of
  deformation modes and the presence of $\theta$ modes,
  respectively. More specifically, they studied the case with
  $b_1(T)=0$, and if $b_1(T)\ne 0$ the associated Wilson line modulini
  must also be carefully studied. The first known examples of compact
  rigid associative submanifolds of compact $G_2$ manifolds were given
  in \cite{Corti:2012kd}, the form of their superpotential corrections
  (modulo potential complications from Wilson line modulini)
  were studied in \cite{Halverson:2014tya}. 

  Finally, note that the instanton prefactor $A$ may not have any
  dependence on $G_2$ moduli that violates the axion shift symmetries
  of M-theory, and therefore it seems that any possible dependence must be
  exponential in the moduli. It is often stated that the prefactor
  must be a moduli independent constant due to the shift symmetries,
  but we believe that this claim cannot be consistent with M-theory
  lifts of IIa configurations with chiral matter prefactors
  \cite{Blumenhagen:2006xt,Florea:2006si,Ibanez:2006da} for Euclidean
  D2-brane instantons. In that context the prefactors are chiral
  supermultiplets, not constants, and we see no reason that such
  configurations should be absent in M-theory. Furthermore, the
  complex scalars in these chiral multiplets have vacuum expectation
  values that depend on open string moduli that are lifted in M-theory
  to $G_2$ moduli; these could be turned into $G_2$-moduli dependent 
  constant prefactors by Higgsing the gauge group that charges the
  supermultiplets. We see no reason that such IIa configurations should
  be absent in M-theory lifts, and conclude only that any moduli dependence
  must be consistent with axionic shift symmetries, as in the case of
  exponential dependence.

\vspace{.5cm}
A four-dimensional string arises from wrapping an M5-brane on a
four-cycle $\tilde T = n_i \tilde T_i$, and its worldsheet in
$\bR^{3,1}$ couples to a two-form $B$ which is the four-dimensional
Hodge dual of an axion. This is straightforward to
understand since the string coupling is
\begin{equation}
\int_{WS\times n_i\tilde T_i}\, C_6 = n_i \int_{WS} B^i
\end{equation}
where $WS$ is the worldsheet of the string. If $T$ is coassociative,
then the string is BPS.

On the other hand, considering just simple Lagrangian quantum field
theories rather than compactifications of string theory or M-theory,
Abrikosov-Nielsen-Olesen (ANO) vortex strings exist in vacua where
some number of $U(1)$ symmetries are spontaneously broken. The
simplest case in field theory is given by the abelian Higgs model. This
theory has $G=U(1)$, but the vacuum spontaneously breaks the symmetry
to a subgroup $H=\emptyset$; it is completely broken. The defects in
the theory are determined by the homotopy groups $\pi_i(G/H) =
\pi_i(U(1))$. Thus, since $\pi_1(U(1))=\bZ$ the theory exhibits ANO
vortex strings.  The tension of the critical or BPS ANO vortex string is 
\begin{equation}
T_{ANO} = 2\pi |v|^2 n,
\end{equation}
where $v$ is the vacuum expectation
value of the charged scalar field which performs the breaking and $n$
is the winding number of the vortex. There are also the semi-local
strings \cite{Vachaspati:1991dz} associated with $U(1)$ symmetry
breaking in theories with flat directions, which also has a tension
that scales as $T \sim |v|^2$.

\vspace{.2cm} Finally, a domain wall arises on four-dimensions from
wrapping an M5-brane on a three-cycle $T = n^i T^i$. We do not have
much to say about these at this point, except that BPS domain walls
may exist in four-dimensional theories. See for example
\cite{Acharya:2001dz}.

\subsection{Two Physical Singular Limits}
\label{sec:sing lim}

Having prepared some preliminaries, let us study two types of singular limits.

One singular limit we will study, which we have already mentioned, is when an
associative $T$ or coassociative $\tilde T$ submanifold goes to zero
volume. Since
\begin{equation}
vol(T) = \int_T \Phi = \phi_i \, \int_T \Phi_i > 0
\end{equation}
this gives a moduli-dependent inequality that is violated in the
singular limit where $T$ vanishes, and is therefore similar to a K\"
ahler cone condition in Calabi-Yau manifolds. A similar statement
holds for coassociative submanifolds.  The other singular limit we
will study derives from an inequality due to Joyce that holds
for any non-zero two-form class on a $G_2$ manifold; this makes it
particularly relevant for vacua on Coulomb branches. We will see that,
although the limit itself is not physically natural, this study
suggests the existence of a calibrated submanifold which would be
physically important; in fact, submanifolds in this class may
be ``gauge theory worldvolumes'' in certain cases.

\subsubsection*{Boundaries from a Lemma of Joyce and Gauge Theory
  Worldvolumes}
\label{sec:Joyce proposition inequality}

We begin with the singular limit associated to the Joyce lemma.  Since
non-trivial two-forms and two-cycles determine the structure of
abelian gauge fields and charged particle states in four-dimensions,
it would be useful for understanding gauge enhancement to have a
$G_2$-moduli dependent condition which involves $H^2(X)$ or $H_2(X)$
in some
way. 

We recall from section~\ref{sec:boundary} a lemma of Joyce (which we
called there the ``topological condition'' on $\Phi$):
let $X$ be a $G_2$ manifold with $G_2$-form $\Phi$ and associated
class $[\Phi] \in H^3(X,\bR)$. Then for any non-zero class $[\sigma]
\in H^2(X,\bR)$ we have $[\sigma] \cup [\sigma] \cup [\Phi] < 0$. For
simplicity we will write the condition as
\begin{equation}
  \label{eq:Joyce condition}
  \int_X \sigma \wedge \sigma \wedge \Phi < 0
\end{equation}
for a representative $\sigma$ of $[\sigma]$. Of course, we can
restrict to integral cohomology, taking $[\sigma] \in
H^2(X,\bZ)$, and there is a dual five-cycle
$[\tilde \Sigma] \in H_5(X,\bZ)$, i.e.,
\begin{equation}
\eta(\tilde\Sigma) = \int \sigma \wedge \eta
\end{equation}
for all $\eta\in H^5(X,\mathbb{Z})$.
 However, (\ref{eq:Joyce condition})
also implies that $[-\sigma \wedge \sigma]$ is non-trivial\footnote{We
  introduce the sign so that the inequality may in some cases be a
  positivity condition on an associative cycle.} in
$H^4(X,\bZ)$, and therefore by Poincar\'e duality we also
have\footnote{Since $\sigma$ is non-trivial then so is $\star_\Phi \sigma$
  and $\star_\Phi (-\sigma \wedge \sigma)$, and accordingly by dimension
  counting there must also exist an integral two-cycle $\Sigma$ and
  four-cycle $\tilde D_\Sigma$. However, unlike for $\tilde \Sigma$
  and $D_\Sigma$, there does not exist a canonical map $\sigma
  \rightarrow \Sigma$ or $\sigma \rightarrow \tilde D_\Sigma$; this is
  because $\star_\Phi$ introduces the metric and does not have a canonical
  action on integral (co)homology. Therefore, we focus on 
  $\tilde \Sigma$ and $D_\Sigma$.} a three-cycle $[D_\Sigma]$ dual to
$[-\sigma \wedge \sigma]$, i.e.,
\begin{equation}
\xi(D_\Sigma) = -\int \sigma\wedge\sigma\wedge \xi
\end{equation}
for all $\xi\in H^3(X,\mathbb{Z})$.  Thus, to any two-form $[\sigma] \in
H^2(X,\bZ)$ we have associated five-cycle \emph{and} three-cycle classes
\begin{equation}
  [\tilde\Sigma] \in H_5(X,\bZ) \qquad \text{and} \qquad
  [D_\Sigma]\in H_3(X,\bZ). 
\end{equation}
In the following we will try to consistently use bracketed expressions to
denote homology classes, while those expressions without the brackets will
denote submanifold (perhaps calibrated) representatives of those classes;
e.g. $[D_\Sigma]$ will be a three-cycle class, but $D_\Sigma$ will be
an associative representative of that class.

This is a remarkable condition!  It says that for \emph{any} $U(1)$
symmetry determined by some $\sigma \in H^2(X,\bZ)$ in a $G_2$
compactification of M-theory, we have a canonical map not only to the
expected non-trivial five-cycle $\tilde \Sigma$ via duality with
homology, but also a map to a non-trivial three-cycle $D_\Sigma$ via
duality with homology and the Joyce lemma. In general, however, there
is no requirement that either of these classes have representatives
that are submanifolds, let alone calibrated ones.

Though in section \ref{sec:gauge enhancement} we will present physical
arguments that this is the case under some circumstances, we simply
assume it for now. Then M2-branes and M5-branes can be wrapped on the
representatives $\tilde \Sigma$ and $D_\Sigma$, giving defects in four
dimensions, which are a
\begin{align}
  \text{Magnetic Monopole:}& \qquad \text{from an M5-brane on } \tilde\Sigma \nonumber \\
  \text{Spacetime Instanton:}& \qquad \text{from an M2-brane on } D_\Sigma\nonumber \\
  \text{Domain Wall:}& \qquad \text{from an M5-brane on } D_\Sigma, 
  \label{eq: Joyce proposition associated defects}
\end{align}
where the Joyce lemma ensures the existence of the class $[D_\Sigma]$,
but the assumed positivity is necessary for the existence of the
instanton and the domain wall.

In addition to its role in the existence of these defects, there are other
physical consequences of the lemma (\ref{eq:Joyce condition}). Again 
assuming (for now) that $[D_\Sigma]$ has an associative representative $D_\Sigma$, note
that (\ref{eq:Joyce condition}) can be rewritten
\begin{equation}
0 < \int_{D_\Sigma} \Phi = vol(D_\Sigma) 
\end{equation} 
in which case the lemma can also be interpreted as a condition on
calibrated cycles;
when it is violated, the volume of an associative
submanifold goes to zero. 
As such, the
generic physical lessons from conditions of the second type (which we
will discuss shortly) also apply here.

What is the physical meaning of $vol(D_\Sigma)$?  Since $D_\Sigma$ is related
to a $U(1)$ gauge theory, it is natural that $vol(D_\Sigma)$ might determine
a parameter in the $U(1)$ theory, and it is reasonable to guess that
it is related to the gauge coupling.  If this is true,
the limit in which (\ref{eq:Joyce
  condition}) is violated should be a limit in the gauge coupling. In
fact this can be seen directly: we can write
\begin{equation}
  Vol(D_\Sigma) = \int_X \sigma \wedge \sigma \wedge \Phi = - \int_X \sigma \wedge \star_\Phi \sigma
\end{equation}
using the identity $\star_\Phi \sigma = -\sigma \wedge \Phi$, and from
section \ref{sec:G2 review}
we recall that this expression appears in the four-dimensional gauge
coupling $\frac{1}{g_{YM}^2}$ via Kaluza-Klein reduction of
eleven-dimensional supergravity. Specifically, in the case $b_2(X)=1$ we have
\begin{equation}
  \frac{1}{4g_{YM}^2}=-\frac{\pi}{\l_{11}^3} \int_X \sigma \wedge \star_\Phi \sigma=\frac{\pi}{l_{11}^3} \int_X \sigma \wedge \sigma \wedge \Phi= \pi \frac{vol(D_\Sigma)}{l_{11}^3} .
\end{equation}
and so we see $g_{YM}^2 = \frac{l_{11}^3}{4\pi vol(D_\Sigma)}$.
This is also a result expected
from considering M-theory lifts of IIa compactifications with
spacetime filling D6-branes: there the four-dimensional /gauge
coupling is determined by the volume of the supersymmetric three-cycle
wrapped by the D6-brane, which lifts to a supersymmetric three-cycle
--- an associative submanifold --- in M-theory. From this and (more importantly)
the general argument we conclude
\begin{center}
  \emph{violation of the Joyce lemma is the limit $g_{YM}^2\rightarrow \infty$},
\end{center}
which is not the physical limit we'd like to consider to control gauge
enhancement. We emphasize that this conclusion holds regardless of whether the class
$[D_\Sigma]$ has an associative representative.

Instead, the physical lesson we would like to draw is that for every
$U(1)$ in a $G_2$ compactification, the Joyce lemma implies the existence of
a non-trivial three-cycle $[D_\Sigma]$. If
$[D_\Sigma]$ has an appropriate associative representative $D_\Sigma$,
then the volume of this associative determines the $G_2$ moduli
dependence of the four-dimensional gauge coupling. This result
matches the expectation from dimensional reduction of a
seven-dimensional gauge theory on $\bR^{3,1} \times D_\Sigma$.
Moreover, an associative submanifold is expected to play the role of
gauge theory worldvolume in the singular limit, as we will discuss.  

\vspace{.5cm}

When an associative submanifold $D_\Sigma$ exists, wrapping an
M2-brane on it gives a spacetime instanton with associated classical
suppression factor
\begin{equation}
  e^{-Vol(D_\Sigma)} \sim e^{-\frac{1}{g_{YM}^2}}.
\end{equation}
This instanton may correct the superpotential in some models depending
on the structure of instanton zero modes.  It is initially confusing
that such an effect would appear, since it has the dependence of a
gauge instanton but $U(1)$ theories on their own do not exhibit such
effects. One possible resolution of this puzzle is that such effects
can appear in abelian theories if they arise as broken phases of
non-abelian theories; see section \ref{sec: adjoint gauge enhancement}
for further discussion.

\begin{table}
\centering
\begin{tabular}{c|c|c|c|}
Cycle & Cycle Dimension & M2 & M5 \\ \hline
$D_\Sigma$ & 3 & spacetime instanton & domain wall \\ \hline
$\tilde \Sigma$ & 5 & --- & magnetic monopole \\ \hline
\end{tabular}
\caption{Any non-trivial two-form $\sigma$ in a $G_2$ manifold
  determines not only a dual five-cycle, but also a related three-cycle. If these have representatives
  that are submanifolds, a variety
  of topological defects in four-dimensions, listed above, can be obtained by wrapping M2-branes
  and M5-branes. }
\label{table:sigma defects}
\end{table}

\subsubsection*{Singularities from Calibrated Submanifolds and
  Vanishing Defects}
\label{sec:calibration inequality}

When the metric moduli of a manifold are varied such that a
non-trivial cycle goes to zero volume, a singularity develops, and
this can be done in $G_2$ moduli for associative and coassociative
submanifolds since they are calibrated. That is, since
\begin{equation}
  \label{eq:calibration}
  Vol(T) = \int_T \Phi \qquad \qquad \text{and} \qquad \qquad Vol(\tilde T) = \int_{\tilde T} \star_\Phi \Phi
\end{equation}
for $T$ or $\tilde T$ any associative or coassociative submanifold,
one simple tunes the moduli such that one of the cycles
vanishes and a singularity develops. 
Note that in many examples
the vanishing of a three-cycle is accompanied by the vanishing of a
two-cycle that it contains; we will discuss the physics of this
phenomenon in sections \ref{sec:gauge enhancement} and
\ref{sec:examples}.

M2- and M5-branes wrapped on associative or coassociative submanifolds can
give rise to BPS instantons, BPS domain walls, and BPS strings, and the energetics of
these defects can be controlled by the calibration forms $\Phi$ and
$\star_\Phi \Phi$. Since in gauge theories the existence of topological
defects is often determined by the properties of spontaneous symmetry
breaking, it is natural to wonder whether the vanishing limit of these
instantons, domain walls, and strings may be related to gauge
enhancement. For example, the vanishing tension limit of an
Abrikosov-Nielsen-Olesen vortex string or semi-local string is a
limit in which the abelian Higgs model is unbroken.

\section{Scenarios for Gauge Enhancement}
\label{sec:gauge enhancement}
In this section we would like to identify
some scenarios in which these ideas may be put to use. 
That is, if M-theory on a particular $G_2$
manifold $X$ describes a phase of a broken gauge theory coupled to
gravity, how might one extract some data of the unbroken gauge theories
in the singular limit?  

A simple case that we begin with is to understand $G_2$
compactifications which describe the Coulomb branch of a non-abelian
theory obtained when an adjoint chiral multiplet receives an
expectation value. We will then describe the physics associated with a
circle full of conifolds, including the Higgs and Coulomb branches
associated with the M-theory lift of the deformation and small
resolution.  We speculate that a deformation of that setup would
produce a theory with electrons and positron localized at different
codimension seven singularities, but we leave the detailed explanation
of such a deformation to future work.

\subsection{Non-Abelian Gauge Theories with Adjoint Chiral Multiplets}
\label{sec: adjoint gauge enhancement}
Consider a four-dimensional $\cN=1$ non-abelian gauge theory with
adjoint chiral multiplets, and suppose that there are flat directions
in the scalar potential along which one (or more) of the adjoint
chiral multiplets can receive an expectation value. If the expectation
values for components of the adjoint chiral are generic, the
non-abelian theory is broken from $G$ to $U(1)^{rk(G)}$. The broken
theory exhibits charged massive W-bosons and massless photons, which
give rise to long range forces. This is a Coulomb branch.

In the $d=4$ $\cN=1$ super-Higgs mechanism a massless vector multiplet
eats a massless chiral multiplet to produce a massive vector
multiplet, and only $dim(G)$-$rk(G)$ chiral multiplets may be eaten in
the breaking $G\mapsto U(1)^{rk(G)}$ since this is the number of vector
multiplets that receive a mass.  In the case of a single
adjoint chiral multiplet all but $rk(G)$ of its
components are eaten in the breaking. The $rk(G)$ leftover components
must be uncharged under $U(1)^{rk(G)}$; these are the $rk(G)$ Cartan
elements of the adjoint.  More massless chiral multiplets remain
(on the Coulomb branch) if there are more adjoint chirals in the
non-abelian theory, or other matter fields.

Such a theory exhibits 't Hooft-Polyakov monopoles. These were first
shown to exist \cite{'tHooft:1974qc,Polyakov:1974ek} in a model of
Georgi and Glashow \cite{Georgi:1972cj} where $G=SU(2)$ is broken to
$H=U(1)$ by an adjoint scalar field. The existence of the monopole
solution is guaranteed by the non-trivial homotopy $\pi_2(G/H)=\bZ$,
which in this case can be seen since $G/H$ is topologically an $S^2$,
or alternatively since $\pi_2(G/H)=\pi_1(H)=\bZ$. The first equality
holds by virtue of a long exact sequence
\begin{equation}
\dots \longrightarrow \pi_2(G) \longrightarrow \pi_2(G/H) \longrightarrow \pi_1(H)\longrightarrow \pi_1(G)\longrightarrow \dots
\end{equation}
and the facts that $\pi_2(G)=\pi_1(G)=0$ for $G=SU(2)$. 
This situation generalizes easily: for any breaking of
$G\mapsto H=U(1)^{rk(G)}$ for a Lie group $G$ with
$\pi_2(G)=\pi_1(G)=0$ we have 
\begin{equation}
\pi_2(G/H)=\pi_1(H)=\bZ^{rk(G)}
\end{equation}
and the broken theory exhibits magnetic monopoles; these include the
cases where $G=SU(N), Sp(N)$, or $G$ an exceptional simple Lie group,
but not $G=SO(N>2)$ since $\pi_1(SO(N>2))=\bZ_2$. Even in this case
$\pi_2(G/H)$ must be non-trivial. This can be seen using the exact
sequence with $G=SO(N>2)$ and $H=U(1)^{rk(G)}$
\begin{equation}
0\rightarrow \pi_2(G/H) \xrightarrow{f} \bZ^{rk(G)} \xrightarrow{g} \bZ_2 \rightarrow \dots
\end{equation}
where if $\pi_2(G/H)$ were trivial, then $g$ would be injective, which
cannot be true. Thus $\pi(G/H)$ is non-trivial for $G$ a simple Lie
group and $H=U(1)^{rk(G)}$, so these theories contain monopoles.

For $G$ a simple Lie group $\pi_3(G)\ne 0$ and the theory exhibits
Yang-Mills (henceforth, gauge) instantons. It is common to think of 
these topologically non-trivial gauge field configurations in unbroken non-abelian gauge theories, but it is also natural
to consider what happens to them if a charged scalar obtains an expectation
value that breaks $G$. This is an important question in the context of M-theory,
since gauge instantons exist in singular $G_2$ compactifications with non-abelian
gauge symmetry, but if they also exist in the Higgs regime then they may arise geometrically
from the structure of a smooth $G_2$ manifold or a submanifold thereof.

In gauge theory this question was answered by 't Hooft
\cite{'tHooft:1976fv}, who showed that gauge instantons associated to a gauge group
$G$ contribute to the effective action even if $G$ is spontaneously
broken to an abelian or trivial group $H$; see \cite{Csaki:1998vv} for
a general treatment, including for a variety of unbroken subgroups
$H$. The usual BPST instanton solution \cite{Belavin:1975fg} can be
used to construct $SU(2)$ instantons for $SU(2)$ subgroups of
$G$. Suppose a Higgs field $h$ receives an expectation value $\langle
h \rangle=v$.  In the case $v\ne 0$ the instanton size $\rho$ is no
longer a modulus and therefore the finite size instanton does not
solve the equations of motion in the Higgs regime. Nevertheless their
contributions to the effective action may be computed (for example,
using the formalism of constrained instantons \cite{Affleck:1980mp})
and the standard suppression factor $e^{-1/g^2_{YM}}$ still
appears. In the original example of \cite{'tHooft:1976fv} the extra
term in the instanton action induced by $v\ne 0$ is
\begin{equation}
\Delta S = 2\pi^2 |v|^2 \rho^2
\end{equation}
where the numerical prefactor is model dependent. The $|v|^2 \rho^2$
dependence is more general, however, as it follows directly from the
Higgs kinetic term $|D_\mu h|^2\supset|A_\mu h|^2$, as clearly
presented in \cite{Csaki:1998vv}, for example. It will be important
for us that the small ($\rho=0$) instanton of the four-dimensional
gauge theory solves the equations of
motion even in the Higgs regime!

Before addressing this issue in M-theory, it is useful to consider the same
question in a type IIa compactification with a stack of $N$ D6-branes wrapped on a special
Lagrangian submanifold $\cM_3$. Reducing to four dimensions gives a four-dimensional
$U(N)$ gauge theory with gauge coupling depending on $vol(\cM_3)$. The small instanton
of this four-dimensional gauge theory is a brane within a brane \cite{Douglas:1995bn},
which in this case is a Euclidean D2-brane that is wrapped on $\cM_3$. The contribution
to the effective action of this ED2 arises as $e^{-1/g_{YM}^2}$ by the relationship between
the gauge coupling and $vol(\cM_3)$. Now, if $\cM_3$ is in a family of special Lagrangians $\cM_{3,t}$ then there 
is an adjoint chiral multiplet that breaks the $U(N)$ theory to $U(1)^N$ by spreading out the
N D6-branes so that they wrap $\cM_{3,t_i}$ where $i=1,\dots, N$. The non-abelian theory
has been broken, but nevertheless an ED2 on any member of this family is a solution and 
still contributes to the effective action as $e^{-1/g^2_{YM}}$ by virtue of the fact that
every member of the family has the same volume. Whether this instanton contributes to the superpotential
or some other aspect of the effective $\cN=1$ action depends on details of zero modes, but it is a solution
in any case, as predicted by 't Hooft's gauge theory calculation \cite{'tHooft:1976fv}. Lifting this
configuration to M-theory should preserve the volume dependence of the ED2-instanton in the broken theory as the
volume of an associative submanifold.

\vspace{.5cm} Now we turn to the more general question of interest: if M-theory on
a smooth $G_2$ manifold $X$ were to realize such a phase of a gauge
theory, accounting for many or all of these interesting features, how
would one tell from the topology of $X$, and how might one take a
singular limit in which the non-abelian gauge symmetry is restored?
One would like to identify the presence of charged massive W-bosons,
adjoint chiral multiplets, 't Hooft-Polyakov monopoles\footnote{More
  specifically, we will look for monopoles that are associated with
  symmetry breaking in some way. We will have nothing to say about the
  field profiles characteristic of 't Hooft-Polyakov monopoles, but
  will focus instead on the dependence of M-theory monopoles on gauge
  theory parameters, finding that they have a dependence characteristic
  of 't Hooft-Polyakov monopoles.},
and gauge instantons.  For simplicity we assume that $rk(G)=b_2(X)$;
one could also apply the argument below to a $rk(G)$-dimensional
subset of the two-forms, five-cycles, and other associated topological
data.

The $\cN=1$ supersymmetric Georgi-Glashow model is, itself, a good
starting point: suppose M-theory on a $G_2$ manifold $X$ realized a
phase of a gauge theory with $G=SU(2)$ broken to $H=U(1)$ by the
expectation value of single adjoint chiral multiplet $X_a$,
$a\in\{1,2,3\}$, which is the only charged chiral multiplet in the
unbroken theory. A D-flat direction exists for the adjoint chiral by
virtue of the fact that $X_a X_b \delta_{ab}$ is a gauge invariant holomorphic
function. The associated branch of moduli space is a Coulomb branch,
which exhibits massive charged
W-bosons, 't Hooft-Polyakov monopoles, gauge instantons and a finite
gauge coupling.  This theory is
an $\cN=1$ avatar of pure Seiberg-Witten theory, which also exhibits
the above physical effects on its Coulomb branch.

If M-theory on $X$ gives a vacuum on the Coulomb branch of this model,
the existence of only this $U(1)$ requires $b_2(X)=1$, and we have
associated
homology classes
$[\tilde \Sigma] \in H_5(X)$ and $[\Sigma] \in H_2(X)$. In a general
abelian theory $[\tilde \Sigma]$ and $[\Sigma]$ need not satisfy any
other conditions. However, the Georgi-Glashow model is not a general
abelian theory; it is obtained by spontaneous symmetry breaking from a
non-abelian theory, and it exhibits both massive W-bosons and
monopoles. Kaluza-Klein reduction of $d=11$ supergravity on $X$ does
not account for this physics and therefore it should arise from
another source. Wrapped M2-branes and M5-branes provide such a source,
as discussed in section \ref{sec:defects}. Doing so, however, requires
that the classes $[\tilde \Sigma]$ and $[\Sigma]$ have representatives
that are positive volume submanifolds; we call these $\tilde \Sigma$
and $\Sigma$. Thus, the physics of the Coulomb branch requires that $X$
satisfy certain properties if the massive W-bosons and monopoles are to
arise from wrapped M2-branes and M5-branes.

What about the gauge instantons? They are naturally present if the 
three-cycle ensured by the Joyce lemma has a representative that
is an associative submanifold. Recall that this lemma gives
\begin{equation}
  [D_\Sigma] \equiv - [\tilde \Sigma] \cap [\tilde \Sigma] \in H_3(X,\bZ),
\end{equation}
a non-trivial three-cycle associated to the $U(1)$, and also that the
gauge coupling is
\begin{equation}
g^2_{YM}=\frac{l_{11}^3}{-4\pi\int_X \sigma \wedge \sigma \wedge \Phi}
%\stackrel{!}{=}
=
\frac{l_{11}^3}{4\pi \, vol(D_\Sigma)} 
\end{equation}
where $\sigma$ is the harmonic two-form that defines the $U(1)$ and the
latter equality holds only if there is an associative submanifold
$D_\Sigma$ in the class $[D_\Sigma]$. If $D_\Sigma$ does exist, then 
M2-branes wrapped on $D_\Sigma$ are spacetime instantons with
$e^{-1/g_{YM}^2}$ type dependence characteristic of gauge instantons.
Since (for example) the small instanton\footnote{The M-theory lift of
  a D6-ED2 system would yield such an configuration, where the ED2 can
  be interpreted as the small instanton of the D6 worldvolume theory
  \cite{Douglas:1995bn}.} does solve the equations of motion in the
Higgs regime, these effects should be visible on the Coulomb branch,
that is in M-theory on $X$!  Furthermore, since they do not arise from
supergravity reduction, an M2-brane instanton on $D_\Sigma$ would give
one of the required $e^{-1/g^2_{YM}}$ suppressed contributions to the
effective action, and the non-triviality of the class $[D_\Sigma]$ is
ensured on general grounds by the Joyce lemma, it seems reasonable
that an associative representative $D_\Sigma$ exists. Henceforth we
assume the existence of such a cycle.

So far, the existence of massive W-bosons, 't
Hooft-Polyakov monopoles, and gauge instantons have given us
submanifold representatives of the classes $[\Sigma]$, $[\tilde
\Sigma]$, and $[D_\Sigma] = -[\tilde \Sigma] \cdot [\tilde \Sigma]$
that are submanifolds. This additional structure on $X$ is not
required in general, but arises from physics. It is interesting to ask
whether physics suggests a relation between the physical parameters in
the Higgs vacuum and the geometry of these submanifolds, in particular
their volumes. Using the standard dependences of W-boson and monopole
masses on the Higgs field expectation value $|v|$ and the gauge
coupling $g_{YM}$ and relating this scaling to the expected scaling of
these the masses with the volumes of $\Sigma$ and $\tilde \Sigma$, we
have
\begin{equation}
M_W \propto g_{YM} |v| \propto vol(\Sigma) \qquad \text{and} \qquad M_M\propto \frac{|v|}{g_{YM}} \propto vol(\tilde \Sigma)
\end{equation}
and therefore the non-abelian limit with $|v| \mapsto 0$ sends both $\Sigma$
and $\tilde \Sigma$ to zero volume, but for the gauge coupling to remain finite $vol(D_\Sigma)$ must
also remain finite. 

The  reader may have already noticed that (in a certain sense) we have an
overconstrained system: the volumes of the three submanifolds $\Sigma$, $\tilde \Sigma$
and $D_\Sigma$ scale with two physical parameters $v$ and $g_{YM}$. Since $\frac{1}{g^2_{YM}}\propto vol(D_\Sigma)$
we can write another scaling for the monopole mass
\begin{equation}
M_M\propto \frac{|v|}{g_{YM}} \propto \frac{M_W}{g^2_{YM}} \propto vol(\tilde \Sigma) \propto vol(\Sigma)vol(D_\Sigma)
\end{equation}
so that the volume of the five-manifold $\tilde \Sigma$ depends on the volumes of the two-manifold
$\Sigma$ and the three-manifold $D_\Sigma$. 

While perhaps not absolutely necessary, this dependence is
\emph{suggestive} of a fibration! Specifically, a fibration $\tilde
\Sigma \rightarrow D_\Sigma$ with generic fibers of class
$[\Sigma]$. If it were true that $\tilde \Sigma=\Sigma \times D_\Sigma$
then we would have $vol(\tilde \Sigma) = vol(\Sigma)\, vol(D_\Sigma)$,
but this would also be true if equivalent volume two-manifolds of
class $[\Sigma]$ were fibered over $D_\Sigma$.\footnote{The latter
  situation is precisely what occurs (with appropriate dimension
  changes for $\tilde \Sigma$ and $D_\Sigma$) in the M-theory Coulomb
  branch obtained from compactification on a resolved Calabi-Yau
  elliptic fibration, which is often used as a tool to study F-theory;
  see appendix \ref{sec:comparison to ECY} for a discussion.} Though
the possibility of a fibration has arisen in our analysis from rather
general properties of $G_2$ manifolds and Coulomb branches, this
structure matches other known ideas in the literature. In non-compact
models a singular ALE space (with ADE singularity) is fibered over a
three-manifold that we might as well call $D_\Sigma$; alternatively
one could get an ADE singularity fibered over $D_\Sigma$ from models
with heterotic duals, in which $X$ fibered over $D_\Sigma$ by
coassociative K3 manifolds. In either case, the resolution of the
singularity would produce five-manifolds that are fibered over
three-manifolds by two-spheres, which in our language would be $\tilde
\Sigma$, $D_\Sigma$ and fibers of class $[\Sigma]$.  If there was such
a fibration, then in the gauge enhanced singular limit the fiber must
collapse and therefore $\tilde \Sigma$ collapses to the three-manifold
$D_\Sigma$. The enhanced non-abelian gauge symmetry is localized on
the three-manifold $D_\Sigma$ over which the singularity has
developed.

Now, $D_\Sigma$ may have its own topology. If $b_1(D_\Sigma)\ne 0$
it has one-cycles, and the $S^2$-fibration over the one-cycles give
three-manifolds that are just $S^2$-fibrations over circles. In the
gauge enhanced singular limit where the W-boson mass goes to zero
the three-manifold would collapse to a circle and we would have the
seven-dimensional gauge theory on $D_\Sigma$. What does the topology
$b_1(D_\Sigma)$ correspond to physically?  Reduction of the
seven-dimensional (on $\bR^{3,1}\times D_\Sigma$) gauge fields on the
$b_1(D_\Sigma)$ one-cycles in $D_\Sigma$ give rise to adjoint chiral
supermultiplets in the $d=4$ $\cN=1$ effective theory, where in the
case of the Georgi-Glashow model we must have
$b_1(D_\Sigma)=1$.

\vspace{.5cm} To this point we have given physical arguments for what
must occur in the singular limit, but we have not discussed how one
gets there. In a Calabi-Yau compactification (again c.f. appendix
\ref{sec:comparison to ECY}) one would could simply calibrate the
fibers to zero, since they are a family of holomorphic curves.  Since
this cannot be done in $G_2$ manifolds, let us use our proposal to use
associative or coassociative submanifolds to develop singularities. 
Recall that the $D_\Sigma$ ensured by the Joyce lemma is not the one
to calibrate to zero, though, since this would be the infinite
coupling limit.

However, physical considerations lead to another three-cycle, namely the $S^2$ fibration over
the one-cycle in $D_\Sigma$ corresponding to the adjoint chiral. Since
in the gauge enhanced singular limit the $S^2$ must go to zero volume,
this seems to be the natural cycle to try to send to zero. Doing so
via calibration requires an associative submanifold $T$ in that class,
so that our proposal is to take a limit in $G_2$ moduli such that 
\begin{equation}
vol(T) = \int_T \Phi \qquad \mapsto \qquad 0,
\end{equation}
where the collapse of $T$ to an $S^1$ seems most natural in order
to retain a finite gauge coupling. 

\vspace{.5cm}
The proposal put forth here in the Georgi-Glashow model extends to Coulomb
branches associated to other non-abelian groups $G$, as well, with some
small modifications. For example, in cases with $rk(G)>1$ there are a wider
variety of monopoles and therefore there must be additional five-manifolds
on which to wrap M5-branes; this, too, matches the Calabi-Yau case discussed 
in appendix \ref{sec:comparison to ECY}, since higher rank groups give rise
to additional Cartan divisors in the resolution. For gauge coupling unification in the singular limit,
it seems that distinct five-cycles $\tilde \Sigma_1$ and $\tilde \Sigma_2$ associated
the monopoles of the Coulomb branch must have
associated associative submanifold $D_{\Sigma_1}$ and $D_{\Sigma_2}$ in the same class,
i.e. $[D_{\Sigma_1}]=[D_{\Sigma_2}]$. The relationship between
the geometry and the gauge coupling, W-boson mass, and monopole mass also generalizes.

\subsection{Massless Charged Matter from Circles of Conifolds}

In this section we will utilize a similar idea --- where two-cycles are
controlled since they sit in calibrated three-cycles --- in theories that
realize conifold transitions, following \cite{Halverson:2014tya}. 
Recall first the physics of an M-theory
compactification on a Calabi-Yau threefold with conifold
singularities, focusing for now on one of the conifolds. It gives rise
to a $U(1)$ gauge theory with massless charged particles, as can be
understood by taking a limit of the resolved conifold. In the resolved
conifold, the singular tip has been replaced by a two-sphere, and an
M2-brane wrapped on the two-sphere gives rise to a massive charged
particle. The K\" ahler form calibrates holomorphic curves in Calabi-Yau
manifolds, and thus upon taking a limit in K\" ahler moduli space the
curve shrinks, giving a massless charged hypermultiplet localized
at a codimension six singularity. When the conifold is deformed this
charged hypermultiplet receives an expectation value, the two-form
which gave rise to the $U(1)$ factor no longer exists and the theory
is spontaneously broken. 

Similar singularities may exist at codimension six in $G_2$
compactifications of M-theory, and though there is no calibration for
two-cycles, the proposal from section \ref{sec:topological defects and
  singular limits} is to control them indirectly by other
calibrations. We begin from the resolution and will address
the deformation in the next section. Consider a smooth $G_2$
manifold $X$ where one can identify a circle full of resolved
conifolds. Compact $G_2$ manifolds with such a feature are known to
exist, as studied for example in \cite{Halverson:2014tya}, and such a
geometry locally appears as
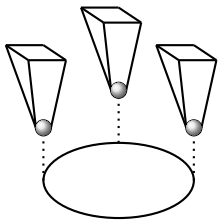
\begin{center}
  \begin{tikzpicture}
    \begin{scope}[shift={(-10mm,3mm)}]
    \draw[thick] (0,4mm) circle (.1cm); 
    \draw[thick,dotted] (0,-3mm)--(0,3mm);
    \draw[thick] (-1mm,4mm)-- (-2mm,13mm);
    \draw[thick] (1mm,4mm)-- (3mm,13mm);
    \draw[thick] (-2mm,13mm) -- (3mm,13mm);
    \draw[thick] (-5mm,15mm) -- (-2mm,13mm);
    \draw[thick] (-5mm,15mm) -- (0mm,15mm);
    \draw[thick] (0mm,15mm) -- (3mm,13mm);
    \draw[thick] (-1mm,4mm)-- (-5mm,15mm);
    \shade[ball color=black!10!white,opacity=0.80] (0,4mm) circle (.1cm);
    \end{scope}
    \begin{scope}[shift={(0mm,8mm)}]
    \draw[thick] (0,4mm) circle (.1cm); 
    \draw[thick,dotted] (0,-3mm)--(0,3mm);
    \draw[thick] (-1mm,4mm)-- (-2mm,13mm);
    \draw[thick] (1mm,4mm)-- (3mm,13mm);
    \draw[thick] (-2mm,13mm) -- (3mm,13mm);
    \draw[thick] (-5mm,15mm) -- (-2mm,13mm);
    \draw[thick] (-5mm,15mm) -- (0mm,15mm);
    \draw[thick] (0mm,15mm) -- (3mm,13mm);
    \draw[thick] (-1mm,4mm)-- (-5mm,15mm);
    \shade[ball color=black!10!white,opacity=0.80] (0,4mm) circle (.1cm);
    \end{scope}
    \begin{scope}[shift={(10mm,3mm)}]
    \draw[thick] (0,4mm) circle (.1cm); 
    \draw[thick,dotted] (0,-3mm)--(0,3mm);
    \draw[thick] (-1mm,4mm)-- (-2mm,13mm);
    \draw[thick] (1mm,4mm)-- (3mm,13mm);
    \draw[thick] (-2mm,13mm) -- (3mm,13mm);
    \draw[thick] (-5mm,15mm) -- (-2mm,13mm);
    \draw[thick] (-5mm,15mm) -- (0mm,15mm);
    \draw[thick] (0mm,15mm) -- (3mm,13mm);
    \draw[thick] (-1mm,4mm)-- (-5mm,15mm);
    \shade[ball color=black!10!white,opacity=0.80] (0,4mm) circle (.1cm);
    \end{scope}
    \draw[thick] (0mm,0mm) ellipse (10mm and 5mm);
  \end{tikzpicture}
\end{center}
where the cone object is the six-dimensional resolved conifold, a
two-sphere has replaced the singular tip as usual and in fact the
conifold is fibered over a circle. Though we cannot control the
two-spheres directly, note that the two-spheres fibered over the
circle give a three-cycle $[T]\in H_3(X,\bZ)$ which contains the
two-spheres, and we consider an associative representative $T$ of this
class. If we move in moduli space such that
\begin{equation}
Vol(T)=\int_T \Phi \qquad \mapsto \qquad 0
\end{equation}
there are at least two natural ways in which the three-manifold $T$
might collapse, analogous to how a divisor may collapse to a curve or
a point in an algebraic variety.  First, $T$ may collapse to a point
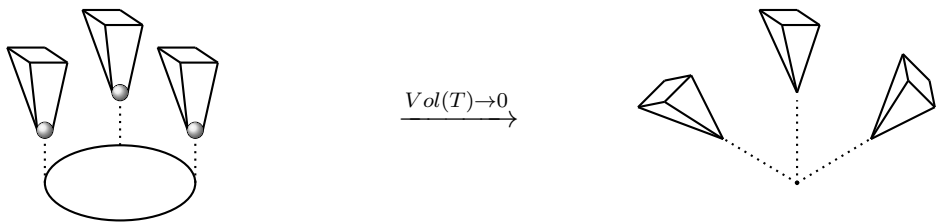
\begin{center}
  \begin{tikzpicture}
    \begin{scope}[shift={(-10mm,3mm)}]
    \draw[thick] (0,4mm) circle (.1cm); 
    \draw[thick,dotted] (0,-3mm)--(0,3mm);
    \draw[thick] (-1mm,4mm)-- (-2mm,13mm);
    \draw[thick] (1mm,4mm)-- (3mm,13mm);
    \draw[thick] (-2mm,13mm) -- (3mm,13mm);
    \draw[thick] (-5mm,15mm) -- (-2mm,13mm);
    \draw[thick] (-5mm,15mm) -- (0mm,15mm);
    \draw[thick] (0mm,15mm) -- (3mm,13mm);
    \draw[thick] (-1mm,4mm)-- (-5mm,15mm);
    \shade[ball color=black!10!white,opacity=0.80] (0,4mm) circle (.1cm);
    \end{scope}
    \begin{scope}[shift={(0mm,8mm)}]
    \draw[thick] (0,4mm) circle (.1cm); 
    \draw[thick,dotted] (0,-3mm)--(0,3mm);
    \draw[thick] (-1mm,4mm)-- (-2mm,13mm);
    \draw[thick] (1mm,4mm)-- (3mm,13mm);
    \draw[thick] (-2mm,13mm) -- (3mm,13mm);
    \draw[thick] (-5mm,15mm) -- (-2mm,13mm);
    \draw[thick] (-5mm,15mm) -- (0mm,15mm);
    \draw[thick] (0mm,15mm) -- (3mm,13mm);
    \draw[thick] (-1mm,4mm)-- (-5mm,15mm);
    \shade[ball color=black!10!white,opacity=0.80] (0,4mm) circle (.1cm);
    \end{scope}
    \begin{scope}[shift={(10mm,3mm)}]
    \draw[thick] (0,4mm) circle (.1cm); 
    \draw[thick,dotted] (0,-3mm)--(0,3mm);
    \draw[thick] (-1mm,4mm)-- (-2mm,13mm);
    \draw[thick] (1mm,4mm)-- (3mm,13mm);
    \draw[thick] (-2mm,13mm) -- (3mm,13mm);
    \draw[thick] (-5mm,15mm) -- (-2mm,13mm);
    \draw[thick] (-5mm,15mm) -- (0mm,15mm);
    \draw[thick] (0mm,15mm) -- (3mm,13mm);
    \draw[thick] (-1mm,4mm)-- (-5mm,15mm);
    \shade[ball color=black!10!white,opacity=0.80] (0,4mm) circle (.1cm);
    \end{scope}
    \draw[thick] (0mm,0mm) ellipse (10mm and 5mm);
    \begin{scope}[shift={(80mm,3mm)}]
    \draw[thick,dotted] (10mm,-3mm)--(0,3mm);
    \end{scope}
    \begin{scope}[shift={(83mm,3mm)},rotate=45]
    \draw[thick] (0mm,4mm)-- (-2mm,13mm);
    \draw[thick] (0mm,4mm)-- (3mm,13mm);
    \draw[thick] (-2mm,13mm) -- (3mm,13mm);
    \draw[thick] (-5mm,15mm) -- (-2mm,13mm);
    \draw[thick] (-5mm,15mm) -- (0mm,15mm);
    \draw[thick] (0mm,15mm) -- (3mm,13mm);
    \draw[thick] (0mm,4mm)-- (-5mm,15mm);
    \end{scope}
    \begin{scope}[shift={(90mm,8mm)}]
    \draw[thick,dotted] (0,-8mm)--(0,3mm);
    \draw[thick] (0mm,4mm)-- (-2mm,13mm);
    \draw[thick] (0mm,4mm)-- (3mm,13mm);
    \draw[thick] (-2mm,13mm) -- (3mm,13mm);
    \draw[thick] (-5mm,15mm) -- (-2mm,13mm);
    \draw[thick] (-5mm,15mm) -- (0mm,15mm);
    \draw[thick] (0mm,15mm) -- (3mm,13mm);
    \draw[thick] (0mm,4mm)-- (-5mm,15mm);
    \end{scope}
    \begin{scope}[shift={(100mm,3mm)}]
    \draw[thick,dotted] (-10mm,-3mm)--(0,3mm);
    \end{scope}
    \begin{scope}[shift={(97mm,3mm)},rotate=-45]
    \draw[thick] (0mm,4mm)-- (-2mm,13mm);
    \draw[thick] (0mm,4mm)-- (3mm,13mm);
    \draw[thick] (-2mm,13mm) -- (3mm,13mm);
    \draw[thick] (-5mm,15mm) -- (-2mm,13mm);
    \draw[thick] (-5mm,15mm) -- (0mm,15mm);
    \draw[thick] (0mm,15mm) -- (3mm,13mm);
    \draw[thick] (0mm,4mm)-- (-5mm,15mm);
    \end{scope}
    \fill[thick,xshift=90mm] (0mm,0mm) ellipse (.3mm and .3mm);
    \node at (45mm,10mm) {$\xrightarrow{Vol(T)\rightarrow 0}$};
\end{tikzpicture}
\end{center}
or alternatively $T$ may collapse to a circle
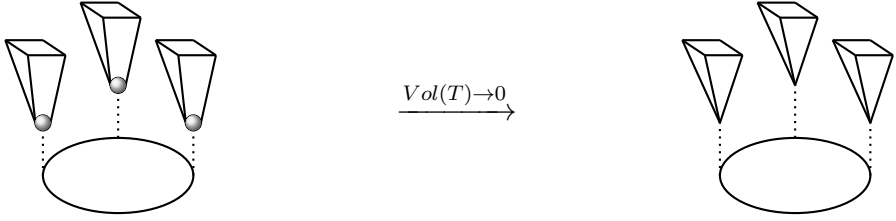
\begin{center}
  \begin{tikzpicture}
    \begin{scope}[shift={(-10mm,3mm)}]
    \draw[thick] (0,4mm) circle (.1cm); 
    \draw[thick,dotted] (0,-3mm)--(0,3mm);
    \draw[thick] (-1mm,4mm)-- (-2mm,13mm);
    \draw[thick] (1mm,4mm)-- (3mm,13mm);
    \draw[thick] (-2mm,13mm) -- (3mm,13mm);
    \draw[thick] (-5mm,15mm) -- (-2mm,13mm);
    \draw[thick] (-5mm,15mm) -- (0mm,15mm);
    \draw[thick] (0mm,15mm) -- (3mm,13mm);
    \draw[thick] (-1mm,4mm)-- (-5mm,15mm);
    \shade[ball color=black!10!white,opacity=0.80] (0,4mm) circle (.1cm);
    \end{scope}
    \begin{scope}[shift={(0mm,8mm)}]
    \draw[thick] (0,4mm) circle (.1cm); 
    \draw[thick,dotted] (0,-3mm)--(0,3mm);
    \draw[thick] (-1mm,4mm)-- (-2mm,13mm);
    \draw[thick] (1mm,4mm)-- (3mm,13mm);
    \draw[thick] (-2mm,13mm) -- (3mm,13mm);
    \draw[thick] (-5mm,15mm) -- (-2mm,13mm);
    \draw[thick] (-5mm,15mm) -- (0mm,15mm);
    \draw[thick] (0mm,15mm) -- (3mm,13mm);
    \draw[thick] (-1mm,4mm)-- (-5mm,15mm);
    \shade[ball color=black!10!white,opacity=0.80] (0,4mm) circle (.1cm);
    \end{scope}
    \begin{scope}[shift={(10mm,3mm)}]
    \draw[thick] (0,4mm) circle (.1cm); 
    \draw[thick,dotted] (0,-3mm)--(0,3mm);
    \draw[thick] (-1mm,4mm)-- (-2mm,13mm);
    \draw[thick] (1mm,4mm)-- (3mm,13mm);
    \draw[thick] (-2mm,13mm) -- (3mm,13mm);
    \draw[thick] (-5mm,15mm) -- (-2mm,13mm);
    \draw[thick] (-5mm,15mm) -- (0mm,15mm);
    \draw[thick] (0mm,15mm) -- (3mm,13mm);
    \draw[thick] (-1mm,4mm)-- (-5mm,15mm);
    \shade[ball color=black!10!white,opacity=0.80] (0,4mm) circle (.1cm);
    \end{scope}
    \draw[thick] (0mm,0mm) ellipse (10mm and 5mm);
    \begin{scope}[shift={(80mm,3mm)}]
    \draw[thick,dotted] (0,-3mm)--(0,4mm);
    \draw[thick] (0mm,4mm)-- (-2mm,13mm);
    \draw[thick] (0mm,4mm)-- (3mm,13mm);
    \draw[thick] (-2mm,13mm) -- (3mm,13mm);
    \draw[thick] (-5mm,15mm) -- (-2mm,13mm);
    \draw[thick] (-5mm,15mm) -- (0mm,15mm);
    \draw[thick] (0mm,15mm) -- (3mm,13mm);
    \draw[thick] (0mm,4mm)-- (-5mm,15mm);
    \end{scope}
    \begin{scope}[shift={(90mm,8mm)}]
    \draw[thick,dotted] (0,-3mm)--(0,4mm);
    \draw[thick] (0mm,4mm)-- (-2mm,13mm);
    \draw[thick] (0mm,4mm)-- (3mm,13mm);
    \draw[thick] (-2mm,13mm) -- (3mm,13mm);
    \draw[thick] (-5mm,15mm) -- (-2mm,13mm);
    \draw[thick] (-5mm,15mm) -- (0mm,15mm);
    \draw[thick] (0mm,15mm) -- (3mm,13mm);
    \draw[thick] (0mm,4mm)-- (-5mm,15mm);
    \end{scope}
    \begin{scope}[shift={(100mm,3mm)}]
    \draw[thick,dotted] (0,-3mm)--(0,4mm);
    \draw[thick] (0mm,4mm)-- (-2mm,13mm);
    \draw[thick] (0mm,4mm)-- (3mm,13mm);
    \draw[thick] (-2mm,13mm) -- (3mm,13mm);
    \draw[thick] (-5mm,15mm) -- (-2mm,13mm);
    \draw[thick] (-5mm,15mm) -- (0mm,15mm);
    \draw[thick] (0mm,15mm) -- (3mm,13mm);
    \draw[thick] (0mm,4mm)-- (-5mm,15mm);
    \end{scope}
    \draw[thick] (90mm,0mm) ellipse (10mm and 5mm);
    \node at (45mm,10mm) {$\xrightarrow{Vol(T)\rightarrow 0}$};
\end{tikzpicture}
\end{center}
which is the case that we will focus on.  In such a limit $X$ exhibits
a circle full of conifolds, and there are massless charged particles
localized at these codimension six singularities. In summary, given a
circle full of resolved conifolds in $X$, the three-cycle $T$ which is
a two-sphere fibration over a circle can be calibrated to zero volume
to control the mass of the charged particles associated with the
two-spheres; this matter is necessarily non-chiral.

Furthermore, in this case (where the associative threefold collapses
to a circle) one can be quite explicit about the local geometry. To do
so, recall that the conifold can be represented by a gauged linear
sigma model (GLSM). Let $(x_1,x_2,x_3,x_4)$ be fields (homogeneous
coordinates) with $U(1)$ charges $(1,1,-1,-1)$ respectively. The
vacuum moduli space of this GLSM solves the D-term constraint
\begin{equation}
|x_1|^2 + |x_2|^2 - |x_3|^2 - |x_4|^2 = \xi,
\label{eq:conifold GLSM D-term}
\end{equation}
where $\xi$ is the Fayet-Iliopoulos (FI) parameter of the GLSM; this
is not to be confused with the FI parameter of the $U(1)$ theory on
spacetime. The vacuum moduli space is a toric Calabi-Yau manifold,
which is a conifold for $\xi = 0 $. There are two
different small resolutions of the conifold; here they are given by
$\xi\ne 0$, where the sign of $\xi$ determines which of the two small
resolutions is realized. Tuning $\xi$ corresponds to movement in the
K\"ahler moduli space of the Calabi-Yau manifold, and knowing this the
$G_2$ interpretation is immediate. In the case of a Calabi-Yau product
with a circle, movement in K\" ahler moduli space induces movement in
$G_2$ moduli space, and thus either of the small resolutions not only
moves in the K\" ahler moduli space but also deforms the $G_2$-form $\Phi$. An
associative submanifold may therefore collapse in the $\xi\mapsto 0$
limit, which we see explicitly here: the cycle $T$ above collapse to a
circle. This phenomenon also occurs in the compact model studied in
\cite{Halverson:2014tya}, which we will review in section \ref{sec:tcs}.

It may be possible to deform these theories by letting
$\xi$ vary along the circle, leaving isolated singularities at the points
where $\xi$ vanishes.  We leave the investigation of such a deformation
to future work.

\subsection{Strings from Deformed Circles of Conifolds}

To this point we have not yet used a BPS defect to understand symmetry
breaking: the monopoles of section \ref{sec: adjoint gauge
  enhancement}, which played a critical role, were not BPS. In this section
we would like to understand symmetry breaking of $U(1)$ theories in terms
of the appearance of BPS strings, and their relation to wrapped branes
on the new cycles of the deformation of conifolds.

Consider the circle of conifolds. From wrapping M2-branes and anti
M2-branes on the two-sphere of the resolved circle of conifolds, 
the conifold limit has two chiral multiplets of massless
charged particles, with opposite charge. Consider the field theoretic
description of this setup. The D-term contribution to the
scalar potential is
\begin{equation}
V_D = \frac{g^2}{2}\, (\xi + \ov \phi_+ \phi_+ - \ov \phi_- \phi_-)^2
\end{equation}
where $\phi_+$ and $\phi_-$ are the complex scalars in the charged
chiral multiplets and $\xi$ is the Fayet-Iliopoulos parameter of the four-dimensional
$U(1)$ theory, which must be
zero at the conifold point since the $U(1)$ is unbroken there. In the super-Higgs
mechanism for $\cN=1$ theories in four dimensions a massless vector multiplet eats
a massless chiral multiplet to become a massive vector multiplet. Since $\xi=0$ at
the conifold point the expectation values
\begin{equation}
\langle \phi_+ \rangle = \langle \phi_- \rangle = v 
\end{equation}
determine a D-flat direction in the scalar potential that Higgses the
$U(1)$, which are a subset of the full space of D-flat directions $\xi
+ \ov \phi_+ \phi_+ - \ov \phi_- \phi_-$, where $\xi$ is
field-dependent.

D-flat directions correspond to holomorphic gauge invariant combinations of the fields,
and this one in particular corresponds to 
\begin{equation}
\phi_+ \phi_-,
\end{equation}
Rewriting the scalar fields in polar coordinates $\phi_i = \rho_i e^{i
  \theta_i}$, the phase of $\phi_+ \phi_-$ is
$e^{i(\theta_++\theta_-)}$, and $\theta_+ + \theta_-$ is uncharged
  under the $U(1)$. The combination $\theta_+ - \theta_-$, on the
  other hand, is the eaten by the photon in the Higgsing process. This field theoretic
  description is natural in M-theory: two chiral multiplets became
  massless in taking the singular limit of the resolved circle of
  conifolds, and in Higgsing the theory by deformation one combination
  of these chiral multiplets is eaten, while the one corresponding to
  the flat direction is left massless. The existence of the new chiral
  multiplet of the flat direction on the Higgs branch corresponds to
  the new contribution to $b_3$ from the $S^3\times S^1$ of the deformed
  circle of conifolds. 

Now wrap an M5-brane on the $S^3\times S^1$. This gives a string in four dimensions
that exists in the deformation, but not the resolution.
Classically, its tension depends on 
\begin{equation}
T \sim R\, vol(S^3)
\end{equation}
where $R$ is the radius of the $S^1$, and since the volume of the
$S^3$ is an order parameter for the symmetry breaking the tension is
$T \sim R \, f(v)$ for some function $f$ of the Higgs expectation
value. This string is charged under the two-form that is the
four-dimensional Hodge dual of the linear combination of axions
$\theta_+ + \theta_-$ associated to the D-flat direction. If the
$S^3\times S^1$ is a coassociative submanifold then this string is BPS.

We see that we have a topological defect -- a BPS string -- that
appears in connection with the breaking of a $U(1)$ gauge symmetry in
a four dimensional theory, and its tension depends on the Higgs
expectation value. This basic phenomenon is known in the breaking
of $U(1)$ theories, of course. Famously, the non-supersymmetric
abelian Higgs model supports a stable, Abrikosov-Nielsen-Olesen (ANO) vortex
string with tension
\begin{equation}
T_{ANO} = 2 \pi |v|^2
\end{equation}
for the critical vortex that satisfies the Bogolmo'nyi bound, which
with slight modifications \cite{Penin:1996si,Gorsky:1999hk} can be
extended to $d=4$ $\cN=1$ theories. There are also the semi-local
strings \cite{Vachaspati:1991dz} associated with $U(1)$ symmetry
breaking in theories with flat directions. In fact, the simplest model
with semi-local strings has two charged scalars, just like this model,
and furthermore the semi-local string tension $T_{SL} \sim |v|^2$.

We leave the detailed study of the type of string we obtained from a
wrapped M5-brane to future work, but since its energetics match the discussed field
theoretic expectations relating string tensions to order parameters for $U(1)$ symmetry breaking
and furthermore it is charged (via $C_6$ reduction) under the two-form
dual to the axion $\theta_+ + \theta_-$ of the flat direction, it
seems a promising harbinger of $U(1)$ symmetry breaking in $G_2$
compactifications of M-theory.  Calibrating it to zero via calibrating
a coassociative to obtain an (abelian) gauge enhanced singular limit
fits with the general proposal we put forth in section
\ref{sec:topological defects and singular limits}.

\section{Examples}
\label{sec:examples}
In this section we would like to study a number of examples, in particular circle
products with Calabi-Yau threefolds, some of the $G_2$ manifolds of Joyce, and
also twisted connected sums.

\subsection{Circle Products with Calabi-Yau Threefolds}
\label{sec:example circle products}
Gauge enhancement is relatively easy to understand if
$X=Z\times S^1$ for $Z$ a Calabi-Yau three-fold. Some of the
physics of gauge enhancement follows directly from the Calabi-Yau
geometry in the usual way, but the circle factor also introduces
additional topological structure that will be important.
But let us first review the relevant mathematical facts.

Consider $X=Z\times S^1$, where $Z$ is a Calabi-Yau threefold. Then
$X$ is a compact seven-manifold with $SU(3)$ holonomy. It is a
manifold with $G_2$-structure, and therefore it has a
$G_2$-form $\Phi$. Letting $\theta$ be an angle
coordinate on $S^1$, $\Phi$ is determined by the K\"
ahler form $J$ and holomorphic three-form $\Omega$ on $Z$
as
\begin{align}
  \Phi &= Re(\Omega) + d\theta \wedge J. 
\end{align}
and similarly, $\star_\Phi \Phi$ also depends on $\Omega$ and $J$.

Though volumes of two-manifolds cannot be easily computed in a general
compact $G_2$ manifold, in this case we can make use of the product
structure of the metric and also the fact that
any $[\Sigma]\in H^2(X,\bZ)$ is also an element of $[\Sigma] \in
H^2(Z,\bZ)$. In particular a holomorphic curve $\Sigma$ in $Z$ is also
a volume minimizing two-cycle in $X$. Therefore,
\begin{equation}
vol(\Sigma)= \int_\Sigma
J = \int_\Sigma \frac{\partial}{\partial \theta} \lefthook \Phi,
\end{equation}
where we have written the last equality to emphasize the
connection between the two-cycle volume and the $G_2$-form $\Phi$.
Note also that to any $\Sigma \in H^2(Z,\bZ)=H^2(X,\bZ)$ we have a
three-cycle $T_\Sigma\in H^3(X,\bZ)$ given by $\Sigma \times S^1$, and
furthermore that any three-cycle in $Z$ is also a three-cycle in $X$.

\vspace{.5cm}
Gauge enhancement upon movement in the moduli space of $X$ can occur
in a few different ways. We will study those cases where gauge
enhancement can be seen via the existence of new massless particles in
the limit where the volume of some two-cycles goes to zero, which is
under control in the case of a circle product for the reasons just
discussed. Since $X$
has $SU(3)$ holonomy, the four-dimensional theory actually has
$\cN=2$ supersymmetry. This theory exhibits BPS particles
and accordingly there is a calibration form for two-manifolds, which is given by
\begin{equation}
\frac{\partial}{\partial \theta} \lefthook \Phi = J.
\end{equation}
One can use this fact to calibrate two-cycles to zero in the usual way.
However, given the relationship to $\Phi$ this may also collapse three-manifolds if
they contain non-trivial two-manifolds, as suggested in the more general proposal we
have put forth in section \ref{sec:topological defects and singular limits}.

Let us consider two natural singular limits of the Calabi-Yau
case, in the language of $X$ and $\Phi$. Let $\Sigma$ be a holomorphic
curve in $Z$ that comes in a real two-parameter family, parameterized
by a curve $C$ of genus $g_C$.  If we have
\begin{equation}
R_{S^1}^6 >> vol(Z).
\end{equation}
then for energy scales above the Kaluza-Klein scale of the $S^1$ but
below that of $Z$ we have an effective five-dimensional theory, and
M2-branes on $\Sigma$ give \cite{Katz:1996ht,Witten:1996qb} a $d=5$
$\cN=1$ massive vector multiplet and $g_C$ adjoint hypermultiplets. These
can be reduced to $d=4$ $\cN=2$ multiplets by compactification on the $S^1$.
From the point of view of $X$, $\Sigma$ comes in a three real
parameter family parameterized by $S^1\times C$ and this family gives a massive
$\cN=2$ vector multiplet in four dimensions.  Taking a limit in the
moduli space of $\Phi$ such that $vol(\Sigma)\rightarrow 0$, the
massive vector multiplet becomes massless and the theory exhibits an
$SU(2)$ gauge enhancement; if multiple curves go to zero volume,
generic ADE gauge groups are possible. In a second scenario, $\Sigma$ could
instead be rigid, in which case M2-branes on it give rise to charged
hypermultiplets in four dimensions; then if $\Sigma$ goes to zero
volume there is no gauge enhancement, but the charged hypermultiplet
becomes massless. If this gives a conifold in $Z$, then we have a circle
of conifolds in $X$. In both of these two cases, from the point of view of
$X$ gauge enhancement occurs due to the collapse of two-cycles inside associative
threefolds, exemplifying our proposal from section \ref{sec:topological defects and singular limits}.

\vspace{.5cm} 

To close this section we would like to remind the reader why
two-cycles were controllable in this case, and speculate as to how
this might generalize.  On general grounds one
might hope that, since the metric is determined by $\Phi$ for 
any $G_2$ manifold, integrating $\int_\Sigma v_\Sigma \lefthook
\Phi$ for some distinguished vector field  $v_\Sigma$ computes the volume of
two-manifold $\Sigma$. This was, in fact, what worked for $X = Z\times
S^1$, where $v_\Sigma = \partial/\partial \theta$ and we note that
$v_\Sigma$ is independent of $\Sigma$. Therefore in this case
$v_\Sigma \lefthook \Phi$ is the same two-form for any $\Sigma$. The
vector $\partial/\partial \theta$ generates the direction normal to
$\Sigma$ in $\Sigma \times S^1$. 

It would be interesting in future work to study this idea in more
general setups, where $v_\Sigma$ may differ for various $\Sigma$. This
is somewhat natural from our proposal to collapse two-manifolds by
collapsing associatives or coassociatives. In the case of an
associative $T$ containing a two-manifold $\Sigma$, there is a natural
vector field $v_\Sigma$ that generates the normal direction to
$\Sigma$ in $T$; the idea would be to study a potential relationship
between $\int_\Sigma v_\Sigma \lefthook \Phi$ and the volume of
$\Sigma$.

\subsection{Joyce's Examples}
\label{sec:examples Joyce}
We now turn to study M-theory on compact seven-manifolds with holonomy
precisely $G_2$. The first examples of these manifolds were due to
Joyce. The idea in this case is to resolve a toroidal orbifold in a
way that allows one to prove that there exists a $G_2$ metric on a
small deformation of the resolved space. Therefore we will denote this
space
\begin{equation}
X = Res(T^7/\Gamma).
\end{equation}
Here $\Gamma$ is some discrete group, which typically acts on the
$T^7$ coordinates in simple ways. For early work on the physics of M-theory
on Joyce orbifolds, see \cite{Acharya:1998pm}.

\vspace{1cm}
We will study two specific examples. The two examples differ
qualitatively: in first, there is only one topologically distinct
smoothing of the singular space, whereas there are two in the second
example. The basic idea is that the singularities can locally be
written as a three-manifold times a real codimension four ADE
singularity, and these singularities can be deformed or resolved into a
smooth four-manifold. Though these four-manifolds are diffeomorphic,
the actions of orbifold elements on them are not necessarily
topologically equivalent. In the first example we study they are, whereas
in the second example they are not.

Before proceeding to the examples, let us discuss some
details of the two desingularizations which are common to both
examples; there the singular sets are locally modeled by $T^3 \times
\mathbb{C}^2/\{ \pm 1\}$ and we focus on one.  Define
$(x_1,x_2, x_3)$ to be the $T^3$ coordinates and $z_1 \equiv x_4 +
ix_5$ and $z_2 \equiv x_6 + i x_7$ to be the coordinates on $\bC^2$,
with the orbifold action locally giving the quotient of $\bC^2$. There
are two possible choices for resolving $\bC^2/\{\pm 1\}$ into a smooth
space $Y_i$:
\begin{itemize}
\item[A)] Let $Y_1$ be the blowup of $\bC^2/\{\pm 1\}$ at the
origin. The exceptional divisor in $Y_1$ is a $\Sigma_1 \cong \bP^1$
and its homology class generates $H_2(Y_1,\bR)$.
\item[B)] Define the map $\sigma: \bC^2/\{\pm 1\} \rightarrow \bC^3$
such that $\sigma: \pm(z_1,z_2) \mapsto (z_1^2 - z_2^2,iz_1^2 +
iz_2^2, 2z_1z_2)$. Then $\bC^2/\{\pm 1\}$ is $\{(w_1,w_2,w_3)\in\bC^3:
w_1^2+w_2^2+w_3^2 = 0 \}$ and the \emph{smoothing} of the singularity
is defined by
\begin{equation} Y_2 \equiv \{ (w_1,w_2,w_3)\in\bC^2:
w_1^2+w_2^2+w_3^2 = \epsilon\} \end{equation} for $\epsilon \in \bC$.
Defining $\epsilon = r e^{2i\theta}$ we have
\begin{equation}\Sigma_2 \equiv \{
(e^{i\theta}x_1,e^{i\theta}x_2,e^{i\theta}x_3): x_j \in \bR, x_1^2 +
x_2^2+ x_3^2 = r\}\end{equation} an $S^2$ in $Y_2$ whose homology
class generates $H_2(Y_2,\bR)$.
\end{itemize} In the respective cases the singular set is now smoothed
into $T^3 \times Y_i$, and $Y_1$ and $Y_2$ are diffeomorphic. The
question in examples will be to study how the orbifold action on each
$Y_i$ could be topologically distinct, and the associated physical
implications.

To fully control this behavior, we would like certain cycles within the blowup
locus to have associative or coassociative representatives, but this is hard
to guarantee.  However, at least in some cases \cite{MR3110581}, there
are $G_2$-instantons whose degeneracy locus approaches the singular locus
as nonabelian gauge symmetry is restored, and we obtain a condition of
the form \eqref{eq:G2-inst}.

\subsubsection*{Example One: A Higgsed $SU(2)$ Theory with a Single
  Branch of Moduli Space}
In this section we will analyze a geometry that determines a moduli
space that spontaneously breaks an $SU(2)^{12}$ gauge symmetry to
$U(1)^{12}$ with three adjoint Higgs fields for each $SU(2)$
factor. We will see that the geometry exhibits topological defects of
the type expected from gauge theoretic arguments, in particular 't
Hooft-Polyakov monopoles, including the correct dependence of monopole
masses on the parameters of the spontaneously broken gauge theory. This
closely matches the discussion of section \ref{sec: adjoint gauge enhancement}.

We begin by reviewing a singular $T^7$ orbifold, discussed for example
in \cite{joyce2000compact}, along with the details that will be
relevant for our analysis.  Let $x_i$ be coordinates on $T^7$ such
that $x_i \sim x_i+1$ and define the actions
\begin{align}
\alpha&: \qquad (x_1,\dots,x_7) \mapsto (x_1,x_2,x_3,-x_4,-x_5,-x_6,-x_7) \nonumber \\
\beta&: \qquad (x_1,\dots,x_7) \mapsto (x_1,-x_2,-x_3,x_4,x_5,\frac{1}{2}-x_6,-x_7)\nonumber \\ 
\gamma&: \qquad (x_1,\dots,x_7) \mapsto (-x_1,x_2,-x_3,x_4,\frac{1}{2}-x_5,x_6,\frac{1}{2}-x_7).
\end{align}
Then $\Gamma = \langle \alpha,\beta,\gamma\rangle$ is the discrete
group that determines the toroidal orbifold. It can be shown that
$\alpha^2 = \beta^2 = \gamma^2=1$ and furthermore that $\alpha,\beta,$
and $\gamma$ commute; as such, $\Gamma \cong \bZ_2 \times \bZ_2 \times
\bZ_2.$ The elements $\beta \gamma$, $\gamma \alpha$, $\alpha \beta$,
and $\alpha \beta \gamma$ do not have fixed points on $T^7$. The fixed
points of $\alpha$ in $T^7$ are $16$ copies of $T^3$, and the group
$\langle \beta, \gamma \rangle$ acts freely on these $16$
$T^3$'s. Similarly, there are another $16$ three-tori fixed by $\beta$
which $\langle \alpha, \gamma \rangle$ acts freely on, and yet another
$16$ three-tori fixed by $\gamma$ which $\langle \alpha, \beta
\rangle$ acts freely on.  The singular set of $T^7/\Gamma$ is $S$,
which is a disjoint union of $12$ three-tori, and the singularity at
each three-torus is locally modeled by $T^3 \times \bC^2/\{\pm
1\}$. Note that the counting $12$, rather than $48$, can be seen by
noting that (for example) the $16$ $T^3$'s fixed by $\alpha$ in $T^7$
are permuted by $\langle \beta, \gamma \rangle$, and thus these are
only $4$ distinct $T^3$'s in the quotient. Similar results hold for
the other fixed $T^3$'s in $T^7$, yielding overall $12$ copies of
$T^3$ in the orbifold. There are no singularities in
codimension seven.

For simplicity, focus one of these singular sets $T^3\times
\bC^2/\{\pm 1\}$.  Desingularizing this space via method A) above, it
is easy to see that that $\Delta b^2(X) = 1$ while $\Delta b^3(X)=3$;
the resolution of the $A_1$ singularity $\bC^2/\{\pm 1\}$ gives the
exceptional curve $\Sigma_1$ that now contributes to $b^2$, while
defining $S_i^1$ with $i\in \{1,2,3\}$ to be the three $S^1$'s in
$T^3$ we see that $S_i^1 \times \Sigma_1$ contributes to $b^3(X)$.  To
match the notation of section \ref{sec:gauge enhancement}, let's
instead use $\Sigma$ for $\Sigma_1$, and we will abuse notation by
letting $\Sigma$ denote either a homology class or a representative,
with meaning determined by context. Let $\tilde \Sigma$ be the
five-cycle that is fibered by two-spheres of class $\Sigma$ over
$T^3$, and we may refer to $T^3$ as $D_\Sigma$. Then M2-branes and
M5-branes on $\Sigma$ and $\tilde \Sigma$ give electrically and
magnetically charged particles in four dimensions, respectively.

\vspace{.5cm} The physics of this example, focusing on this particular
resolved singularity, is as follows. Massive W-bosons arise from
wrapping M2-branes and anti M2-branes on curves of class $\Sigma$
where $SU(2)$ has been broken to $U(1)$; the latter follows since
$\tilde \Sigma$ has a dual two-form $\sigma$ and therefore gives rise
to a massless Z-boson via $C_3 = A\wedge
\sigma + \cdots$. The theory on $\bR^{3,1}\times D_\Sigma$ has a gauge field,
and therefore adjoint chiral multiplets can be obtained by
Kaluza-Klein reduction of the gauge field on one-cycles in
$D_\Sigma$.  In addition to the W-bosons, we can obtain other charged
particles by wrapping an M5-brane on $\tilde \Sigma$, which are
magnetically charged. Note from the generic analysis of section
\ref{sec:G2 review} and also direct dimensional reduction of the gauge
kinetic term in the $d=7$ theory on $\bR^{3,1}\times D_\Sigma$, we
have $Vol(D_\Sigma)\sim 1/g^2$, where $g$ is the four-dimensional
gauge coupling. Noting also that $Vol(\tilde \Sigma)=Vol(\Sigma)\times
Vol(D_\Sigma)$ and the fact that $M_W \sim g |v| \sim Vol(\Sigma)$
with $v$ a Higgs expectation value, we see that the mass $M_m$ of an
M5-brane on $\tilde \Sigma$ is
\begin{equation}
  M_m^2 \sim vol(\tilde \Sigma) \sim vol(\Sigma) vol(T^3) \sim g |v| \times \frac{1}{g^2} \sim \frac{|v|}{g}
\end{equation}
which is precisely the dependence on $v$ and $g$ expected for the classical mass
of a critical 't Hooft-Polyakov monopole. The physical content in this example
closely matches our more general discussion (see section \ref{sec: adjoint
  gauge enhancement}) of Coulomb branches.

Gauge enhancement occurs in this example as discussed in general in
section \ref{sec: adjoint gauge enhancement}: associated to each $S_i^1$ in
$D_\Sigma$ is an adjoint chiral multiplet and a vertical three-cycle
$T_i$ which is a two-sphere fibration over the $S_i^1$. The singular limit is
the Joyce orbifold itself, in which the volume of $T_i$ is approximated 
by\footnote{We thank Thomas Walpuski for comments on this point.
Finding a way to control this approximation will be an interesting topic
for future work.}
\begin{equation}
  vol(T_i) \sim \int_{T_i} \Phi \mapsto 0
\end{equation} 
and since $D_\Sigma$ remains of finite volume in the orbifold limit we
know that the two-cycle $\Sigma$ also vanishes in that limit. Joyce's
work shows that in fact all three $T_i$ vanish, and therefore we have
a vanishing two-cycle $\Sigma$ over every point in $D_\Sigma$. These
give rise to massless W-bosons and the theory in the singular limit
exhibits $SU(2)$ gauge symmetry for each $D_\Sigma$.

\subsubsection*{Example Two: Two Topologically Distinct Resolutions}
Let's study an explicit example; further mathematical details can be
found in section $12.3$ of \cite{joyce2000compact}. Let $x_i$ be
coordinates on $T^7$ such that $x_i \sim x_i+1$. Define the actions
\begin{align}
\alpha&: \qquad (x_1,\dots,x_7) \mapsto (x_1,x_2,x_3,-x_4,-x_5,-x_6,-x_7) \nonumber \\
\beta&: \qquad (x_1,\dots,x_7) \mapsto (x_1,-x_2,-x_3,x_4,x_5,\frac{1}{2}-x_6,-x_7)\nonumber \\ 
\gamma&: \qquad (x_1,\dots,x_7) \mapsto (-x_1,x_2,-x_3,x_4,-x_5,x_6,\frac{1}{2}-x_7).
\end{align} Defining $\Gamma \equiv \langle \alpha,\beta,\gamma
\rangle \cong \mathbb{Z}_2^3$, we are interested in the orbifold
$T^7/\Gamma$. Important facts are that $\alpha$, $\beta$, and $\gamma$
commute, that $\alpha^2 = \beta^2 = \gamma^2 = 1$, and that
$\beta\gamma$, $\gamma\alpha$, $\alpha \beta$, and $\alpha \beta
\gamma$ have no fixed point in $T^7$. The fixed points of $\alpha$,
$\beta$, and $\gamma$ are each $16$ copies of $T^3$.  The set of $T^3$
in $T^7$ fixed by $\alpha$ and $\beta$ are acted on freely by the
groups $\langle \beta, \gamma \rangle$ and $\langle \alpha, \gamma
\rangle$, respectively; the set of $T^3$ in $T^7$ fixed by $\gamma$
are not acted on freely by $\langle \alpha, \beta \rangle$, however,
since $\alpha\beta$ will map some of the $T^3$ to themselves, albeit
without fixed points.  The singular set $S$ of $T^7/\Gamma$ is a
disjoint union of $8$ copies of $T^3$ and $8$ copies of
$T^3/\mathbb{Z}_2$; the former arise from the quotient action on the
32 $T^3$ in $T^7$ fixed by $\alpha$ and $\beta$, whereas the latter
arise from the quotient action on the 16 $T^3$ fixed by $\gamma$.

Codimension four singularities arising along the 8 $T^3/\bZ_2$
singular sets are more exotic. In each of these cases the singularity is
locally modeled by $(T^3 \times \mathbb{C}^2/\{ \pm 1\})/\langle
\alpha\beta \rangle$, where the action of $\alpha\beta$ is given by:
\begin{equation}
\alpha\beta: \left[(x_2,x_4,x_6), \pm(z_1,z_2)\right] \mapsto \left[(-x_2,-x_4,x_6 + \frac{1}{2}), \pm (z_1,-z_2)\right]
\end{equation} where . This corrects a small typo in equation (12.7) of
\cite{joyce2000compact}.  
As in the previous case, the singular sets $T^3\times \bC^2/\{\pm 1\}$ can be resolved in two diffeomorphic ways as $T^3\times Y_i$,
but now we must also take into account the action of $\alpha \beta$ on
the resolution. By the above action, one can consider the induced
action on the homogeneous coordinates of the exceptional divisor
$\Sigma_1$ in $Y_1$; it is such that the orientation of $\Sigma_1$ is
preserved. Alternatively, in the second description $\alpha \beta$
acts as $(w_1,w_2,w_3) \mapsto (w_1, w_2, w_3)$. The induced action on
$\Sigma_2$ is $(x_1,x_2,x_3) \mapsto (x_1,x_2,-x_3)$ which fixes
$\Sigma_2$ but is orientation reversing. Thus, $(\alpha\beta)_*
[\Sigma_1] = [\Sigma_1]$ in $H_2(Y_1,\bR)$ and
$(\alpha\beta)_*[\Sigma_2] = -[\Sigma_2]$ in $H_2(Y_2,\bR)$.  
Thus, for codimension four singularities of this type, resolutions of
type A still contribute a new two-cycle, but resolutions of type B do
not, due to the orientation reversal.  Let $C_j$ be new three-cycles,
with $S^1$ coordinate $x_i$, associated with the resolution of $T^3
\times \bC^2/\{\pm 1\}$. The action of $\alpha\beta$ on the $x_j$ determines
whether the three-cycle $C_j$ is also a three-cycle of the quotient space, and
thus the $G_2$ manifold. For resolutions of type A, $C_6$ is a three-cycle in
the quotient space, whereas $C_2$ and $C_4$ are three-cycles of the quotient for
resolutions of type B. In summary, after quotienting by $\alpha\beta$,
resolutions of type $A$ contribute one new two-cycle and one new three-cycle,
and resolutions of type $B$ contribute zero new two-cycles and two new three-cycles.

We see that for codimension four singular loci of this second type,
the two different types of smoothings give topologically distinct
$G_2$ manifolds! Due to the existence and non-existence of new
two-cycles for cases $A$ and $B$, respectively, one of the smoothings
has a $U(1)$ gauge symmetry and the other does not. We expect that
the singular model will have non-abelian gauge symmetry and sit at
the intersection of a Higgs branch and Coulomb branch. We leave more
detailed study of this model to future work.

\subsection{Twisted Connected Sum $G_2$ Manifolds}
\label{sec:tcs}

In recent years there has been much progress in constructing compact
$G_2$ manifolds using the so-called twisted connected sum (TCS)
construction of Kovalev \cite{KovalevTCS}. The compactification of
M-theory on TCS $G_2$ manifolds was the subject
\cite{Halverson:2014tya}, to which we refer the reader for further
background on the TCS construction. In this section we would like to review the
twisted connected sum construction and discuss how the concrete
example of \cite{Halverson:2014tya} exemplifies our proposal from
section \ref{sec:topological defects and singular limits}. Though it
would require further mathematical developments to realize examples,
we would also like to put forward a new proposal for taking singular
limits of TCS $G_2$ manifolds that realize non-abelian gauge symmetry.

Let us begin by briefly reviewing the twisted connected sum construction,
giving the reader a qualitative understanding and referring the
summary in section $2$ of \cite{Halverson:2014tya} for more details.
In two sentences, the construction glues together two
non-compact seven-manifolds with $G_2$ structure to obtain a compact
seven-manifold $X$. If the non-compact seven-manifolds and gluing satisfy certain
criteria, then $X$ admits a natural $G_2$ structure with associated $G_2$ form $\tilde \Phi$,
and there is a torsion-free deformation $\Phi$ of $\tilde \Phi$ (within its cohomology class),
so that the space $X$ together with the metric $g_\Phi$ associated to $\Phi$ form a compact
seven-manifold with holonomy $G_2$.

The critical details lie in the criteria that the gluing map and the
non-compact seven-manifolds must satisfy. Let $V_\pm$ be an
\emph{asymptotically cylindrical} (ACyl) Calabi-Yau three-fold, which
is a non-compact Calabi-Yau manifold that asymptotes to a Calabi-Yau
cylinder $V_{\pm}^\infty$ at infinity.  A Calabi-Yau cylinder has the
form $\Sigma\times \bC^*$ where $\Sigma$ is a smooth K3 surface and
the $\bC^*$ is often thought of as an interval times $S^1$. The point
is that the Calabi-Yau cylinder, which is a threefold, has a K\"ahler
form and holomorphic three-form inherited from the K\" ahler and
complex structure of $\Sigma$ together with data from the
$\bC^*$. This subsequently induces a natural $G_2$ structure at the
asymptotic ends of the non-compact seven-manifolds $M_\pm := V_\pm
\times S^1_\pm$, and the TCS construction works by gluing the
asymptotic ends of $M_\pm$ in a way such that the compact
seven-manifold $X$ obtained by the gluing has a natural $G_2$
structure and associated $G_2$ form $\tilde \Phi$. To perform such a
gluing requires finding a Donaldson matching, which is a
diffeomorphism $r$ that glues the smooth K3 surfaces $\Sigma_\pm$ in a
way that leaves the $G_2$ structure on the asymptotic ends of $M_\pm$
invariant. Then, Kovalev's theorem showed that there is a torsion free
deformation $\Phi$ of $\tilde \Phi$ in its cohomology class so that
$(X,g_\Phi)$ is a compact $G_2$ manifold.

The construction of compact $G_2$ manifolds using the TCS construction
is then reduced to the construction of ACyl Calabi-Yau threefolds
$V_\pm$ and the study of gluing maps. In some cases the gluing map can
be chosen somewhat trivially, so that the problem is reduced to
constructing the ``building blocks'' $V_\pm$. Practically, this is
often done by constructing associated compact K\" ahler threefolds
$Z_\pm$ with a smooth K3 surface $\Sigma_\pm$ in the anticanonical
class, which is then cut out to form $V_\pm := Z_\pm \setminus
S_\pm$. In \cite{MR3109862} it was shown that such $V_\pm$ can be
constructed from weak Fano threefolds $Z_\pm$, which greatly increased
the number of TCS $G_2$ manifolds to over fifty million \cite{Corti:2012kd}. 
The integral cohomology of TCS $G_2$ manifolds is known, as are some construction
theorems for compact associative submanifolds useful for physics \cite{Corti:2012kd}.

\subsubsection*{Our Proposal and a $U(1)^3$ Example}
In \cite{Halverson:2014tya} we studied a four-dimensional M-theory
compactification on a TCS $G_2$ manifold $X$ that exhibited $U(1)^3$
gauge symmetry, a rich spectrum of massive charged particles, and
(modulo issues of potential Wilson line modulini) instanton
corrections to the superpotential. Furthermore, we studied a non-isolated
conifold transition that broke one of the $U(1)$'s, and Kovalev's theorem
ensures that the this deformation, too, is a compact $G_2$ manifold.

We would like to review this example in just enough detail to make it
clear that it exemplifies the proposal we have outlined in section
\ref{sec:topological defects and singular limits} for taking singular
limits of compact $G_2$ manifolds. For the elements of the physics
that we studied, it sufficed to focus on ``one half'' of $X$, that is,
on one of the seven-manifolds $M= V\times S^1$. The ACyl Calabi-Yau
threefold, originally studied in \cite{MR3109862} but extended with
new results in \cite{Halverson:2014tya}, is obtained as
follows. Consider a one-parameter linear system of divisors --- a
pencil --- in $\bP^3$ generated by K3 surfaces $S_0$ and $S_\infty$ in
the anticanonical class, where $S_\infty$ is a generic quartic and
$S_0 = \{x_1x_2x_3x_4=0 \}$ with $x_i$ the standard homogeneous
coordinates for $\bP^3$. The base locus of this pencil is the union of
curves $C_i$ where each $C_i$ is where $x_i=0$ intersects
$S_\infty$. The threefold $Z$ is obtained by blowing up sequentially
along these curves, and then $V$ is obtained from $Z$ by cutting (the
inverse image of) a generic member of the pencil $|S_0,S_\infty|$.
There are $24$ rigid holomorphic curves in $Z$ that arise from the
blow-up, and these sit away from the locus where $S$ was cut out, so
these are non-trivial in $V$ and also in any $G_2$ manifold $X$ formed
from this building block.

M2-branes wrapped on those $24$ rigid two-cycles give rise to massive
charged particles in four dimensions, and in \cite{Halverson:2014tya}
we studied a limit in which some of these particles become massless,
so that a conifold transition can be performed. Since two-cycles are
not calibrated in a $G_2$ manifold, we took the singular limit by
using the proposal of section \ref{sec:topological defects and
  singular limits}.  Namely, by a theorem of \cite{Corti:2012kd}, to
any rigid holomorphic curve $C$ in $V$ there is a compact rigid
associative in $X$ that is diffeomorphic to $S^2\times S^1$, and the
conifold limit can be taken in $G_2$ moduli by calibrating this
associative to zero volume. Essentially it occurs by continuously
following a family of ACyl Calabi-Yau manifolds $V_t$ to $V_0$ which
has conifold singularities, and Kovalev's theorem can be applied for
any $t\ne 0$ to obtain a $G_2$ manifold. As in the Calabi-Yau case, we
assume that there is a singular Ricci-flat metric even for $t = 0$. In
summary, we obtained a massless particle limit by calibrating a
compact rigid associative submanifold to zero volume, as proposed in
section \ref{sec:topological defects and singular limits}.

\subsubsection*{Towards Non-Abelian Gauge Sectors from K3 Two-cycles}
\label{sec:tcs nags}

How might one take a singular limit of a TCS $G_2$ manifold that
exhibits non-abelian gauge symmetry?  Though the following proposal
would require some addition mathematical progress in order to work,
let us discuss it since it may be a promising direction for future
research. The basic idea is that, since TCS $G_2$ manifolds are
fibered by (not necessarily coassociative) K3 surfaces, one may try to
shrink curves in the K3 surfaces such that a singularity develops in
the TCS $G_2$ manifold $X$.

Recall that one of the ways that we obtained control over particle
masses in the case of the resolved circle of conifolds was that there
was an associative three-cycle that contained a non-trivial two-cycle,
and the limit of vanishing associative led to the collapse of the
two-cycle. One of the reasons that this worked was the fact that
\begin{equation}
H^2(X,\bZ) = K_+ \oplus K_- \oplus\cdots \qquad \qquad H^3(X,\bZ) = K_+ \oplus K_- \oplus \cdots
\end{equation}
where $K_\pm = ker(\rho_\pm)$ with $\rho_\pm: H^2(V_\pm)\to
H^2(S_\pm)$ the natural restriction map. This implies that $K_\pm$
contributes both two-cycles and three-cycles, and it is clear how
they arise: the former are curves $C_\pm \subset V_\pm$ that are
non-trivial in $X$, but taking the product with $S^1_\pm$ in $M_\pm$
also gives three-cycles that are non-trivial in $X$. Moreover, these
three-cycles have an associative representative, and we used that
associative to indirectly control the mass of the M2-branes wrapped on
$C_\pm$. However, $C_\pm$ were rigid, so those M2-branes gave rise to
chiral multiplets, not vector multiplets.

What other non-trivial two-cycles are there in $X$ that might instead give
rise to non-abelian gauge enhancement if we could shrink them via shrinking associatives?
The full second cohomology is
\begin{equation}
H^2(X,\bZ) = K_+ \oplus K_- \oplus N_+ \cap N_-
\end{equation}
where $N_\pm = Im(\rho_\pm)$ and dualizing to homology there may be
two-dimensional submanifold representatives associated to $N_+ \cap
N_-$; these would be divisors in $V_\pm$ that restrict to non-trivial
curves on both of the K3 surfaces $S_\pm$. So, there can be
non-trivial curves in $X$ that are curves in the K3 fibers. M2-branes
on these submanifolds give charged particles
and one may like to study a massless limit of those particles for the
sake of gauge enhancement.  However, since (unlike $K_\pm$) $N_+ \cap
N_-$ does not also contribute to $H^3(X,\bZ)$ or $H^4(X,\bZ)$ it seems
more difficult to shrink those curves via a shrinking associative or
coassociative.  If a singular limit with non-abelian gauge symmetry
exists associated to non-trivial $N_+ \cap N_-$, these would be
Coulomb branches.

Alternatively, one could try to use associatives or coassociatives to
control Higgs branches; that is, to use collapsing associatives or
coassociatives to collapse two-cycles in them that are non-trivial
in the K3 surfaces $S_\pm$ but are trivial in $X$. Doing so would
require that there are curve classes in $S_\pm$ that contribute to the
third or fourth (co) homology but not to the second. But in fact this
is the case! Keeping the pieces of the third and fourth cohomology
that are related to the second cohomology of the K3 surfaces, we have
\cite{Corti:2012kd}
\begin{align}
H^3(X,\bZ) = L/(N_+ \oplus N-) \oplus (N_- \cap T_+) \oplus (N_-\cap T_-) \oplus \cdots \nonumber \\
H^4(X,\bZ) = (T_+ \cap T_-) \oplus L/(N_-\oplus T_+) \oplus L/(N_+ \oplus T_-) \oplus \cdots 
\end{align}
where $H^2(S_\pm)=L$ and 
\begin{equation}
T_\pm = N_\pm^\perp = \{l \in H^2(S_\pm,\bZ) | \langle l,n \rangle=0 \,\, \forall \,\, n \in N_\pm \}.
\end{equation}
To our knowledge, there are currently no theorems related to
associative or coassociative representatives of these
classes. However, if such calibrated submanifolds existed, it is
natural to expect that they contain two-cycles that are non-trivial in
$S_\pm$, but trivial in $X$. If those calibrated submanifolds were to
collapse in a way that the two-cycles collapsed (as in our conifold idea and
examples) then it could develop codimension four singularities in the $K3$ fibers
of $X$, which could be a gauge enhanced singular limit.

Without construction theorems for associative or coassociative
submanifolds related to two-cycles in $S_\pm$ it is difficult to move
forward, but we find this a promising idea that should be pursued in
future work.

\section{Conclusions}
In this paper we have studied gauge enhancement and singular limits in
the compactification of M-theory on a seven-manifold $X$ with holonomy
$G_2$.  Such singular limits are necessary for obtaining a realistic
compactification, since M-theory on $X$ gives rise to at most abelian
gauge symmetry, and therefore cannot realize the standard model of
particle physics. In fact, M-theory on $X$ also cannot give rise to
massless charged particle states.

Our paper develops a proposal and techniques for identifying when
M-theory on a singular limit of $X$ gives rise to massless charged
matter or non-abelian gauge sectors. The proposal and techniques are
motivated by an important fact about $G_2$ manifolds: they do not have
calibrated two-cycles, and therefore it is difficult to track the
volumes of two-cycles as a function of metric moduli. This should be
contrasted to the case of M-theory on a Calabi-Yau manifold, where
calibrated two-cycles are holomorphic curves and their volumes can be
reliably computed as a function of moduli (specifically, K\" ahler
moduli), even though the metric on the Calabi-Yau manifold is not
known.  This is physically important for the following reason.  Since
M2-branes on two-cycles in a smooth manifold give rise to massive
charged particles, the existence of holomorphic curves in Calabi-Yau
manifolds means the W-boson masses can be tracked reliably as a function
of moduli, whereas they cannot in $G_2$ manifolds since the two-cycles
are not calibrated. This is related to the fact that there are
no BPS particles in the $G_2$ compactifications, since it has $\cN=1$
supersymmetry in four dimensions, whereas compactification on a Calabi-Yau
threefold gives an $\cN=1$ theory in \emph{five} dimensions, which does
support BPS particles. Thus, the method of studying non-abelian gauge
enhancement in Calabi-Yau compactifications by sending W-boson masses
to zero via calibrating two-cycles to zero volume cannot be utilized
in a $G_2$ compactification.

In section \ref{sec:topological defects and singular limits} we made a
proposal for studying gauge enhancement and / or singular limits with
massless charged particles in $G_2$ compactifications of M-theory, and
the proposal may be stated both mathematically and
physically. Physically, the proposal is to control harbingers of
symmetry breaking other than the massive W-bosons used in Calabi-Yau
compactifications. Depending on the example, this could
include\footnote{Another signal of symmetry breaking are the 't
  Hooft-Polyakov monopoles, but since these are particles in $d=4$
  $\cN=1$ theories they are not BPS and they are no more useful than
  massive W-bosons; this correlates with the fact that the monopoles
  arise from five-cycles in $X$, which are not calibrated.}  BPS
instantons, strings, or domain walls. Mathematically, this is possible
since these objects arise from M2-branes or M5-branes on calibrated
submanifolds of $X$, which are so-called associative three-manifolds
and coassociative four-manifolds. Therefore, mathematically the
proposal is to study singular limits of M-theory on $X$ by collapsing
associative or coassociative submanifolds. In some cases this may
collapse a two-cycle within the associative or coassociative, in which
case one has gained indirect (in the sense that the two-cycle itself
is not calibrated) control over a particle mass.

In the remainder of the paper we proceeded forward using the proposal,
beginning with general aspects and proceeding toward increasingly
concrete scenarios, finishing with concrete examples. We studied
defects obtained from wrapped M2-branes and M5-branes in section in
\ref{sec:defects} and studied two types of singular limits in section
\ref{sec:sing lim}. One of those limits collapsed a natural cycle that
is explicitly related to symmetry. Namely, for any $[\sigma] \in H^2(X,\bZ)$ a lemma of Joyce shows that
\begin{equation}
[\sigma] \cup [\sigma] \cup [\Phi] < 0
\end{equation}
and therefore $-[\sigma] \cup [\sigma]$ is non-trivial in $H^4(X,\bZ)$
and there is a corresponding three cycle $[D_\Sigma]$ in $H_3(X,\bZ)$.
This is related to a $U(1)$ symmetry via Kaluza-Klein reduction of the
M-theory three-form. However, if $D_\Sigma$ is a calibrated submanifold
representative of class $[D_\Sigma]$, then collapsing $D_\Sigma$ is not
a limit with gauge enhancement, since $vol(D_\Sigma)$ computes the
gauge coupling; instead it is a strongly coupled limit. We mention it, though,
since this class was useful in other ways.

In section \ref{sec:gauge enhancement} we studied some more specific
scenarios for gauge enhancement, in particular Coulomb branches of a
non-abelian gauge theory in section \ref{sec: adjoint gauge
  enhancement} and issues related to massless charged matter, string
defects, and issues related to conifold transitions in the latter
subsections of section \ref{sec:gauge enhancement}.  The Coulomb
branch scenario involved the breaking the $G\to U(1)^{rk(G)}$ by the
expectation values of scalar fields in adjoint chiral
multiplets. Accounting for the 't Hooft-Polyakov monopoles of this
theory with M5-branes wrapped on non-trivial five-manifolds and
considering the relationship between the (classical) monopole masses
and W-boson masses, we were lead to a picture where the five-manifolds
are fibered by two-spheres over a parameter space of class
$[D_\Sigma]$. A three-cycle class related to the adjoint chiral
multiplets appears naturally and is used for gauge enhancement.  This
general picture was exemplified in the first of Joyce's examples
studied in section \ref{sec:examples Joyce}.  We also studied conifold
transitions and saw that the blowdown of the resolution of the circle
of conifolds had a collapsing three-cycle, while the deformation to
the Higgs branch has an emerging four-cycle; these more general
phenomena were exemplified the $U(1)^3$ TCS example discussed in
section \ref{sec:tcs} and studied in \cite{Halverson:2014tya}.

It would be interesting to use the proposal and techniques developed
in this paper for finding new singular limits of $G_2$ compactifications
that exhibit non-abelian gauge symmetry. One possibility is to use our
proposal for twisted connected sum $G_2$ manifolds, perhaps along the lines
discussed in section \ref{sec:tcs nags}. We leave this to future work.

\vspace{.5cm}

\noindent \textbf{Acknowledgments.} We thank Bobby Acharya, Allan
Adams, Csaba Cs\'aki, Mirjam Cveti{\v c}, Keshav Dasgupta, Simon
Donaldson, Antonella Grassi, Tom Hartman, Gordy Kane, Nabil Iqbal,
Albion Lawrence, Andr\'e Lukas, Joe Polchinski, Radu Roiban, Julius
Shaneson, Washington Taylor, and Scott Watson for useful
conversations.  We are particularly grateful to Thomas Walpuski for his
detailed comments on the first version of this paper.
J.H. is grateful to Antonella Grassi and Julius
Shaneson for collaboration on related topics in F-theory, and thanks
J.L. Halverson for her support and encouragement.  We thank the Simons
Center for Geometry and Physics for hospitality and the participants
of the $2014$ Simons Summer workshop for interesting discussions. We
also thank the Aspen Center for Physics for hospitality and support
under NSF grant PHY-1066293. This work is supported by NSF grants
PHY-1307513 and PHY-1125915.

\appendix

\section{Monopoles from Elliptic Calabi-Yau Coulomb Branches}
\label{sec:comparison to ECY}

Since M-theory Coulomb branches arising from Calabi-Yau
compactification have been studied extensively in recent years, in
particular as a tool to study F-theory, we would like to see
how that story parallels Coulomb branches of $G_2$ compactifications.

Consider M-theory compactified on a smooth Calabi-Yau elliptic
fibration $X\xrightarrow{\pi}B$ that is the resolution of a singular
model $\tilde X$ that has ADE singularities along a divisor $Z$ in
$B$. Let us fix $dim_\bC(B)=3$ for concreteness, so that this is a
$\cN=2$ M-theory compactification in $2+1$ dimensions. M-theory on
$\tilde X$ has a non-abelian gauge theory on $Z$ with some gauge group
$G$, and the resolution breaks $G$ to $U(1)^{rk(G)}$ by giving an
expectation value to the adjoint scalar field in the non-abelian
vector multiplet.  Though this discussion generalizes, for simplicity
consider again $G=SU(2)$.
This is an M-theory Coulomb branch from Calabi-Yau compactification,
and though it parallels the $G_2$ case, it also has some properties
that the $G_2$ case doesn't have. First,  geometrically, W-bosons and monopoles of
this theory arise from M2-branes and M5-branes wrapped on curves and
divisors, respectively, which (unlike the two-manifolds and the manifolds in real
codimension two in the $G_2$ case) are calibrated
submanifolds. Second, field theoretically, there are BPS particles in
$d=3$ $\cN=2$ theories.

The resolved geometry has a so-called Cartan divisor $D$, which is a
$\bP^1$ fibration over $Z$, and since $Z$ is a divisor in a threefold
$D$ is itself a threefold. Let the fiber class be $\Sigma$. An
M5-brane wrapped on $Z$ gives rise to a spacetime instanton, which in
$2+1$ dimensions can be a magnetic monopole. This is the correct
interpretation in this case since the M5-brane is a magnetic source
for $C_3$, which gives rise to the abelian vector multiplets of the Coulomb
branch. M2-branes wrapped on the family of curves of class $\Sigma$
give rise to the massive W-bosons of the Coulomb branch, though at
codimension one subloci in $Z$ it is possible that the fiber becomes a
reducible variety, splitting according to $\Sigma = \Sigma_1 +
\Sigma_2$.  Note, though, that the total fiber volume remains the same
even for the split fiber since
\begin{equation}
vol(\Sigma) = \int_\Sigma J = \int_{\Sigma_1} J + \int_{\Sigma_2} J = vol(\Sigma_1) + vol(\Sigma_2).
\end{equation}
The volume of $D$ is
\begin{equation}
vol(D) = \frac{1}{6} \int_D J\wedge J \wedge J = \frac{1}{2} vol(\Sigma) \,vol(Z)
\end{equation}
where the last equality holds due to the fibration.  These volumes
determine the W-boson and monopole mass, respectively, but we haven't
said anything about the gauge coupling yet. In fact, the gauge
coupling depends on $vol(Z)\sim \frac{1}{g^2}$ which here plays a role
analogous to $vol(D_\Sigma)$ in the $G_2$ Coulomb branch. So we see
that the monopole mass, which depends on the volume of the Cartan
divisor $D$, therefore depends on the parameters of the gauge theory as
$|v|/g$ due to $vol(D) \sim vol(\Sigma) vol(Z)$, where $vol(\Sigma)\sim g |v|$
since it determines the W-boson mass.

In summary, this closely matches the $G_2$ Coulomb branch. In both
cases (the $d=4$ $G_2$ Coulomb branch and the $d=3$ Calabi-Yau Coulomb
branch of resolution) there is a real codimension two submanifold that
is a non-trivial cycle. It is fibered by two-spheres over a real
codimension four parameter space, and by wrapping an M5-brane on this
real codimension two submanifold we obtain a magnetic monopole. Since
in both cases the codimension two submanifold arises due to the
smoothing / Higgsing process, the associated monopole is related to
symmetry breaking. This is the behavior expected of 't Hooft-Polyakov
monopoles, and in both cases the monopole mass dependence $|v|/g$
is recovered due to the relationship between geometric volumes and
physical parameters in the broken gauge theory. 

%%%%%%%%%%%%%%%%%%%%%%%%%%%%%%%%%%%%%%%%%%%%%%
%%%%%%%%%%%%%%%%%%%%%%%%%%%%%%%%%%%%%%%%%%%%%%
%%%%%%%%%%%%%%%%%%%%%%%%%%%%%%%%%%%%%%%%%%%%%%
%%%%%%%%%%%%%%%%%%%%%%%%%%%%%%%%%%%%%%%%%%%%%%

\bibliographystyle{JHEP}
\bibliography{refs}

\end{document}